\documentclass[9pt,superscriptaddress,amsmath,amssymb,aps,twocolumn]{revtex4-2}
\usepackage{graphicx}
\usepackage{breakurl}
\usepackage[a4paper, top=2cm, bottom=2cm, left=1.5cm, right=1.5cm]{geometry}
\usepackage[colorlinks=true,linkcolor=blue,urlcolor=blue,citecolor=blue,anchorcolor=blue]{hyperref}

\begin{document}

\title{Simplicial temporal networks from Wi-Fi data in a University Campus: the effects of restrictions on epidemic spreading}
% Controlling epidemics by sideward contact tracing in gatherings
% Epidemics and Sideward Contact Tracing in simplicial temporal networks\\
%Sideward Contact Tracing in simplicial temporal networks\\
%The contribution of Sideward Contact Tracing in gatherings (simplices? clusters? crowds? groups?)\\
%The effectiveness of Sideward Contact Tracing in gatherings\\
%Sideward Contact Tracing in crowds\\
%Tracing simplices to control epidemics}

\author{Andrea Guizzo}
\affiliation{Dipartimento di Scienze Matematiche, Fisiche e Informatiche,
 Universit\`a degli Studi di Parma, Parco Area delle Scienze, 7/A 43124 Parma, Italy}
\affiliation{INFN, Sezione di Milano Bicocca, Gruppo Collegato di Parma, Parco Area delle Scienze, 7/A 43124 Parma, Italy}

\author{Alessandro Vezzani}
\affiliation{Dipartimento di Scienze Matematiche, Fisiche e Informatiche,
 Universit\`a degli Studi di Parma, Parco Area delle Scienze, 7/A 43124 Parma, Italy}
\affiliation{INFN, Sezione di Milano Bicocca, Gruppo Collegato di Parma, Parco Area delle Scienze, 7/A 43124 Parma, Italy}
\affiliation{Istituto dei Materiali per l'Elettronica ed il Magnetismo (IMEM-CNR), Parco Area delle Scienze, 37/A-43124 Parma, Italy}

\author{Andrea Barontini} 
\affiliation{U.O. Sistemi Tecnologici e Infrastrutture, Area Sistemi Informativi, Universit\`a degli Studi di Parma, Parma, Italy}

\author{Fabrizio Russo} 
\affiliation{U.O. Sistemi Tecnologici e Infrastrutture, Area Sistemi Informativi, Universit\`a degli Studi di Parma, Parma, Italy}

\author{Cristiano Valenti} 
\affiliation{U.O. Sistemi Tecnologici e Infrastrutture, Area Sistemi Informativi, Universit\`a degli Studi di Parma, Parma, Italy}

\author{Marco Mamei} 
\affiliation{Dipartimento di Scienze e Metodi dell'Ingegneria, Universit\`a di Modena e Reggio Emilia, Reggio Emilia, Italy}
\affiliation{Artificial Intelligence Research and Innovation Center - AIRI, Universit\`a di Modena e Reggio Emilia, Modena, Italy}

\author{Raffaella Burioni}
%\email{raffaella.burioni@unipr.it}
\affiliation{Dipartimento di Scienze Matematiche, Fisiche e Informatiche,
 Universit\`a degli Studi di Parma, Parco Area delle Scienze, 7/A 43124 Parma, Italy}
\affiliation{INFN, Sezione di Milano Bicocca, Gruppo Collegato di Parma, Parco Area delle Scienze, 7/A 43124 Parma, Italy}


\begin{abstract} % abstract
%\parttitle{First part title} %if any
Wireless networks are commonly used in public spaces, universities and public institutions and provide accurate and easily accessible information to monitor the mobility and behavior of users. Following the application of containment measures during the recent pandemic, we analyse extensive data from the WiFi network in a University Campus in Italy during three periods, corresponding to partial lockdown, partial opening, and almost complete opening. We measure the probability distributions of groups and link activation at Wi-Fi Access Points, investigating how different areas are used in the presence of restrictions. We rank the hotspots and the area they cover according to their crowding and to the probability of link formation, which is the relevant variable in determining potential outbreaks. We consider a recently proposed epidemic model on simplicial temporal networks and we use the measured distributions to infer the change in the reproduction number in the three phases. Our data show that additional measures are necessary to limit the epidemic spreading in the total opening phase, due to the dramatic increase in the number of contacts. 
%\parttitle{Second part title} %if any
%Text for this section.

\end{abstract}

\maketitle

\section{Introduction}

Wireless networks are increasingly used in public spaces as a tool
to provide constant connectivity over large areas for diversified  devices.
Organised in widespread and economic hotspots, wireless networks, in contrast to mobile networks, are particularly accessible to researchers because their use is common in university and corporate campuses and, in general,  in institutions with interest in primary research. Data from wireless networks have often been analyzed to measure the quality of service and to improve network performance and management \cite{Wifi,7577031}. 
However, they also provide accurate and easily accessible data that allow to study users mobility, physical space fruition and users behavior \cite{app1,7447750}.

Monitoring the occupation of public spaces by many people at the same time has become particularly interesting in the last two years, during the COVID-19 pandemic. In fact, in this last period it has become of fundamental importance to be able to control crowded areas in order to maintain adequate distancing measures.  Large events and gatherings, in particular those taking place indoors, have been often related to super-spreading events that have accelerated the 
outbreaks \cite{lau2020}. Therefore they should be monitored. On the other hand, social distancing creates enormous economic and social cost. It is thus important to adequately control large gatherings in public spaces and at the same time quantify the benefits of different containment measures.

In this respect, WiFi data represent an extremely interesting tool. They are easily available by public institutions and they rely on relatively economic infrastructures that are already present in public spaces. The localisation of agents is not as accurate as for GPS and mobile traces \cite{cattuto2010,ISELLA2011,machens2013}, but it can be improved with different tools. Interestingly, Wi-Fi data directly provide the distribution of clusters of users connected to the same access point, potentially crowding the same area and forming temporal simplices which evolve in time \cite{petri2018simplicial,mancastroppa2021sideward}. This is relevant information to be used as an input for models of of higher order interactions in epidemic and information spreading with complex contagion \cite{iacopini19,BATTISTON2020,BATTISTON2021}. 

Considering the problem of pandemic control, in this paper we study WiFi data from the University Campus in Parma, Italy, to monitor both the formation of large gatherings and the presence of areas of intense traffic. We follow the daily usage of public areas across the Campus from anonymized data of devices connected to the access points (APs) of the network. In particular, we measure the probability distribution of large gatherings and of the number of different couples (links) generated at the same spot, potentially accounting for dangerous contacts, in specific areas of the Campus. We analyze three phases, characterized by different containment measures: a closing phase, a partial opening phase and an open phase, with almost complete reopening of the University activities.  Based on this analysis, we rank the APs according to their potential danger and we characterize the fruition of spaces by different classes of users in the three periods.  In particular, we show that the attendance data in the closing phase follow a completely different distribution, while the two partial opening phases share similar features, after a proper rescaling of the total numbers of users. Relying on a recent modelling scheme of epidemic propagation in activity driven simplicial temporal networks \cite{petri2018simplicial,mancastroppa2021sideward,mancastroppa2020quarantine,mancastroppa2021contacttracing}, we also estimate the change in the reproduction number $R_0$ due to the release of restrictions in the last period. Our data signal a dramatic change in the reproduction number, suggesting that additional measures are crucial to limit the probability of outbreaks in the open phase.

\section{WiFi Data}

\subsection{Attendance data in the presence of different containment measures}
The University of Parma, like many universities and public institutions, has covered its buildings and spaces with a unified WiFi network, enabling all users to establish more than 10000 sessions a day.
All sessions data from the login management system are collected by the ``ICT services''  office of the University of Parma. The login management system manages all wireless APs and all users' requests for connection to the internet with their registered devices.

The staff of ``ICT service''  office has the authorization to access files with personal data and carries out an anonymization process. The structure of the dataset and the anonymization process is described in Appendix.
From the anonymized data, we infer the number of individuals and the number of different couples per days in a specific areas, avoiding double-counting of the same individual in a 24-hour time window. These data do not any contain personal information. 

The resulting dataset refers to a sample of 696 wireless APs, 19749 users (divided in 16505 students, 1968 structured staff and 1276 external guests) and about 15000 daily connections. The dataset spans ten months, starting on $10^{TH}$ December $2020$ and ending on $7^{TH}$ November $2021$. During this period, due to the COVID-19 pandemic restrictions, we can distinguish three different phases: a closing phase, a partial opening phase and total opening phase for the $2021/2022$ academic year.

\begin{itemize}
    \item \textbf{Closing phase.} During this phase, the access to  University buildings was allowed only to staff, faculty, and students that took part in laboratory activities. All lectures were held remotely. This phase starts on $10^{TH}$ December $2020$ ends on $21^{TH}$ February $2021$ and starts again on $21^{TH}$ February $2021$ and ends on $18^{TH}$ April $2021$.
    
    \item \textbf{Partially opening phase.} During this phase, the access to the University buildings was extended to first year students to allow them to follow lessons in limited presence (about $25\%$ students enrolled in the degree courses). This phase starts on February $22^{TH}$ 2021, ends on March $14^{TH}$ 2021 and starts again on April $19^{TH}$ 2021, finally ending in June.
    
    \item \textbf{Total opening phase.} During this phase, teaching activities took place in classrooms with the request of a Green Pass to access the University. This phase starts on September $27^{TH}$ 2021 and ends with our dataset.
\end{itemize}

\subsection{Dynamical behavior of attendance data}
\begin{figure*}[]
\centering
\includegraphics[angle = 90, width=0.75\textwidth]{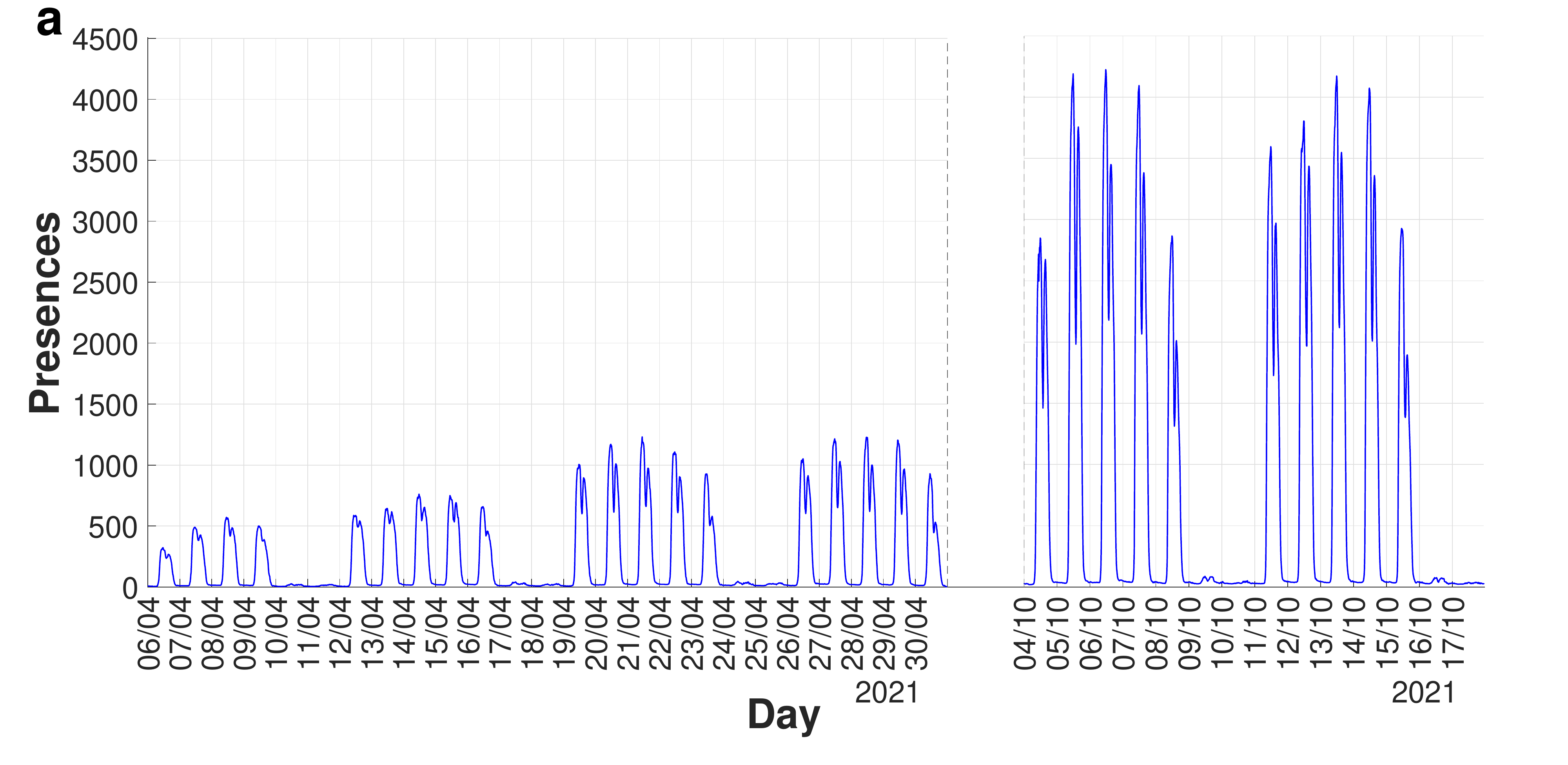}\\
\includegraphics[angle = 90, width=0.75\textwidth]{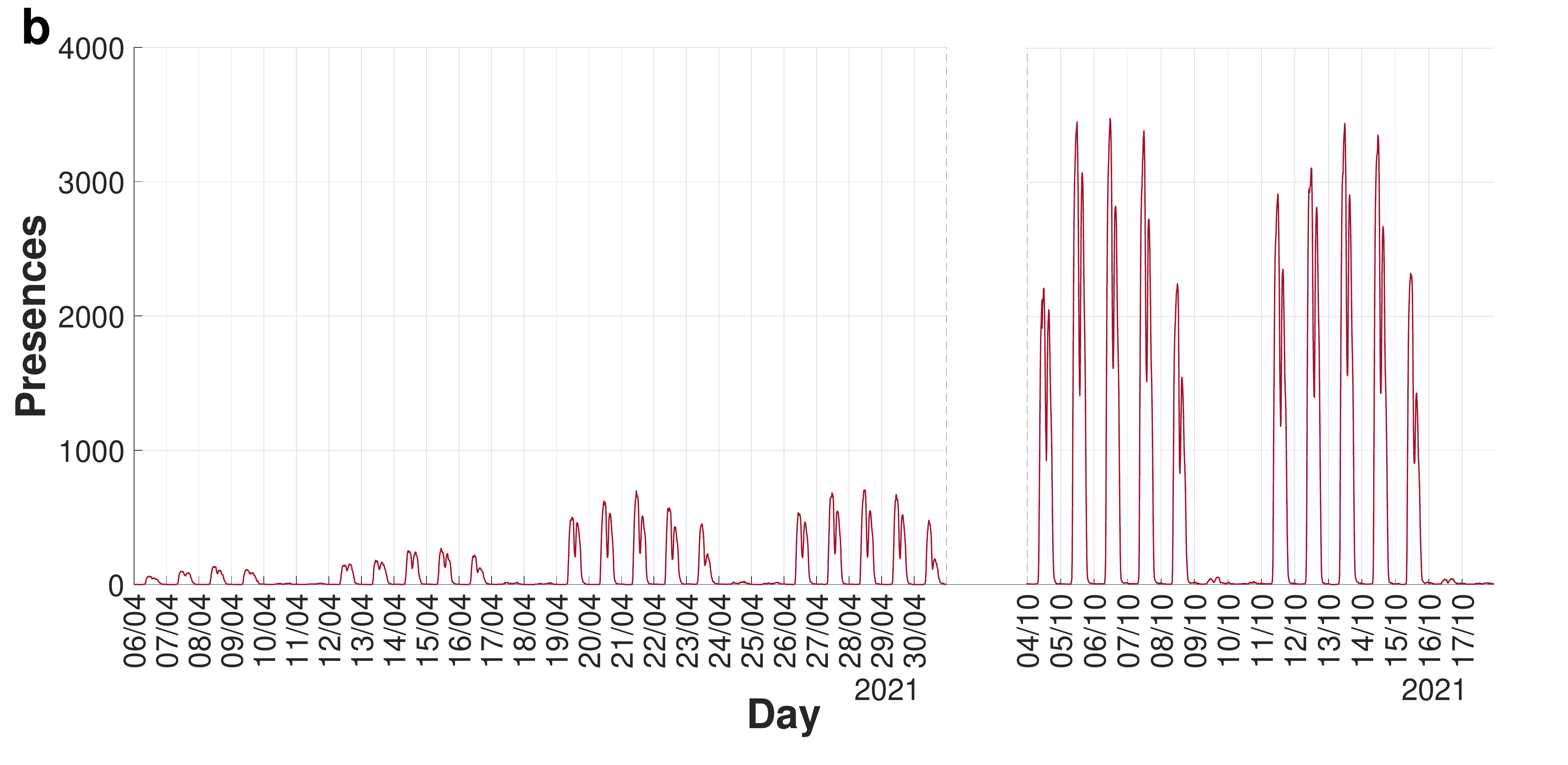}\\
\includegraphics[angle = 90, width=0.75\textwidth]{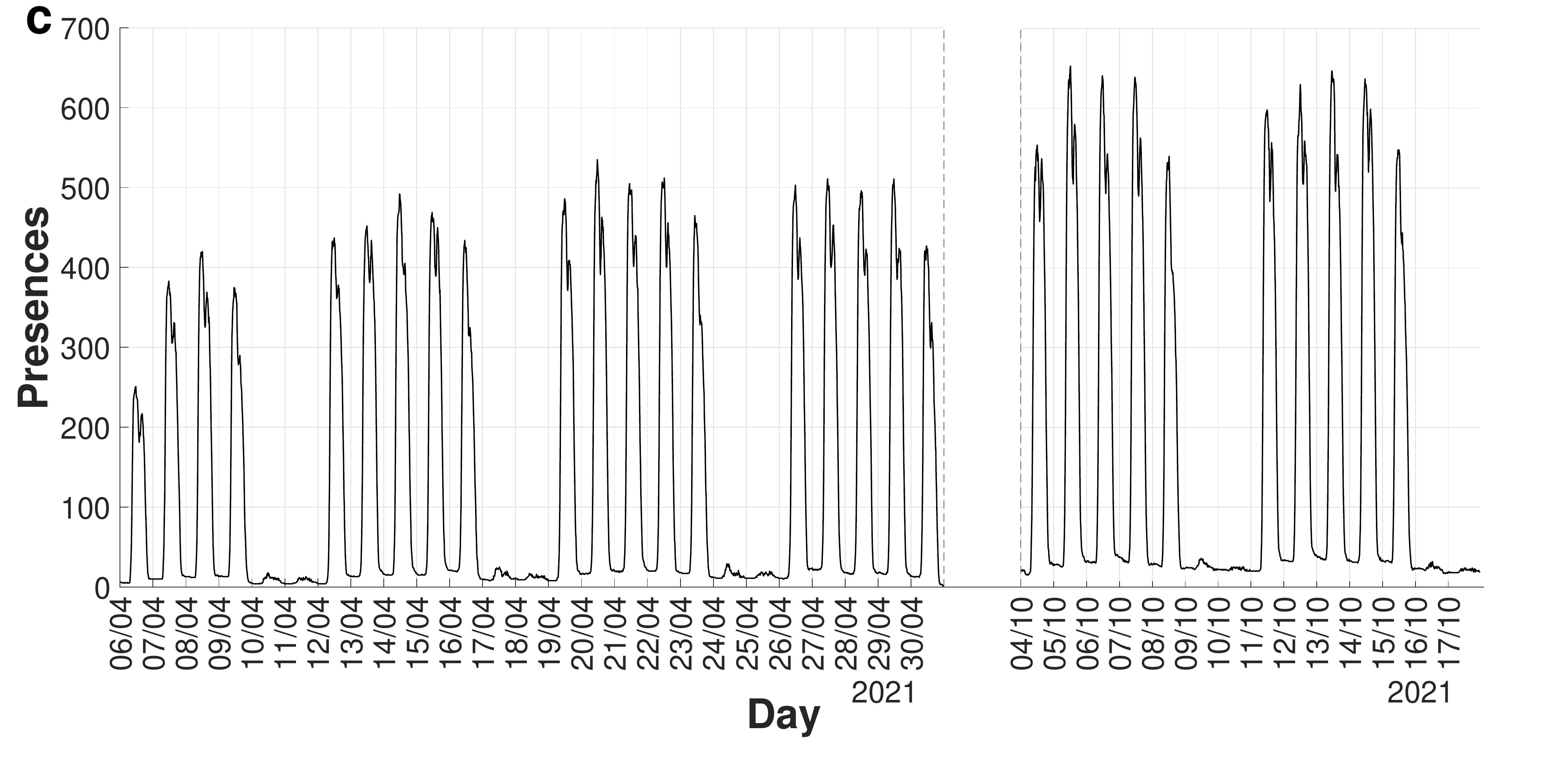}
  \caption{{\bf Dynamical behavior of attendance at Parma University.} Plot of temporal evolution of the attendance of all users (panel {\bf a}), students (panel {\bf b}) and structured staff (panel {\bf c}) during the different closing phases. The first two weeks correspond to the closing phase, the two central weeks to the partial opening phase while the last two weeks to the total opening phase. We can observe a significantly increase in attendance for students in different phases, while for structured staff we find a very similar dynamical behavior of attendance between different regimes. The data are very detailed and give several information, such as the low number of attendances during the weekend and during holidays.
    }
  \label{fig:figure1}
\end{figure*}

From the connection session, we can estimate the time evolution of the attendance at a single AP, in a University building or area (such as the Scientific Campus) or in the whole  University. To obtain these data, we extract from the connection sessions the number of users connected to each AP as a function of time, every minute. The sum of all AP's temporal series corresponds to the presences for the entire University or for the Campus APs. The algorithm also extracts the dynamical behavior of attendance for generic users and, separately, students users, structured staff and external guests. In Fig~\ref{fig:figure1} we graphically represent attendance data at the University in the different phases of closures. For structured staff we find a very limited increase of the attendance in the second and third regimes, while the presence of students increases very significantly in the last phase.

%: this is most likely due the fact that professors and researchers had access to the University buildings and laboratories during the closing period by registering online or by using the badge at the entrance of the University buildings. 
%Probably part of te  researchers and university professors continued their activity in smart working.

\begin{figure*}[]
\centering
\includegraphics[width=0.75\textwidth]{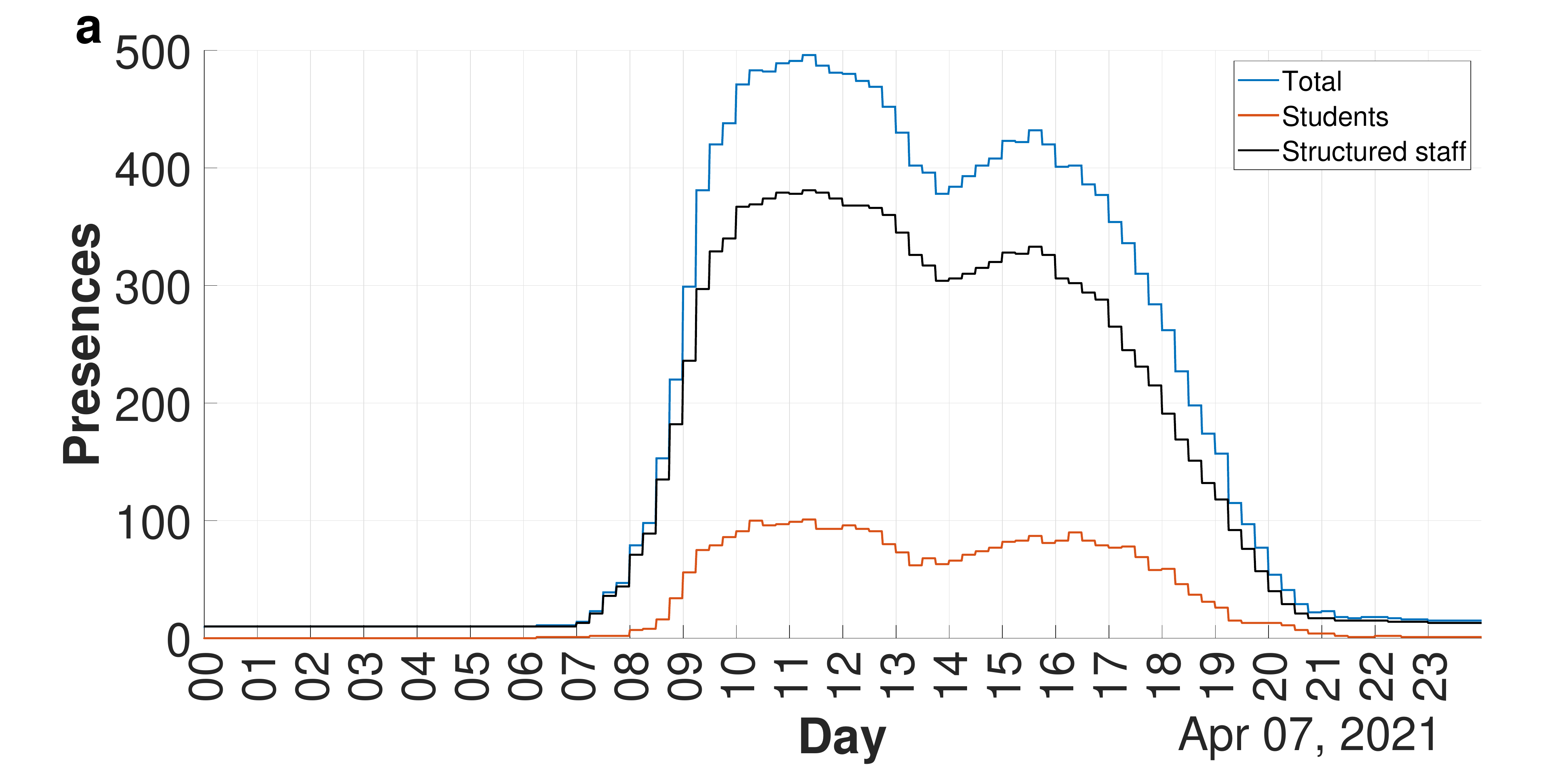}
\includegraphics[width=0.75\textwidth]{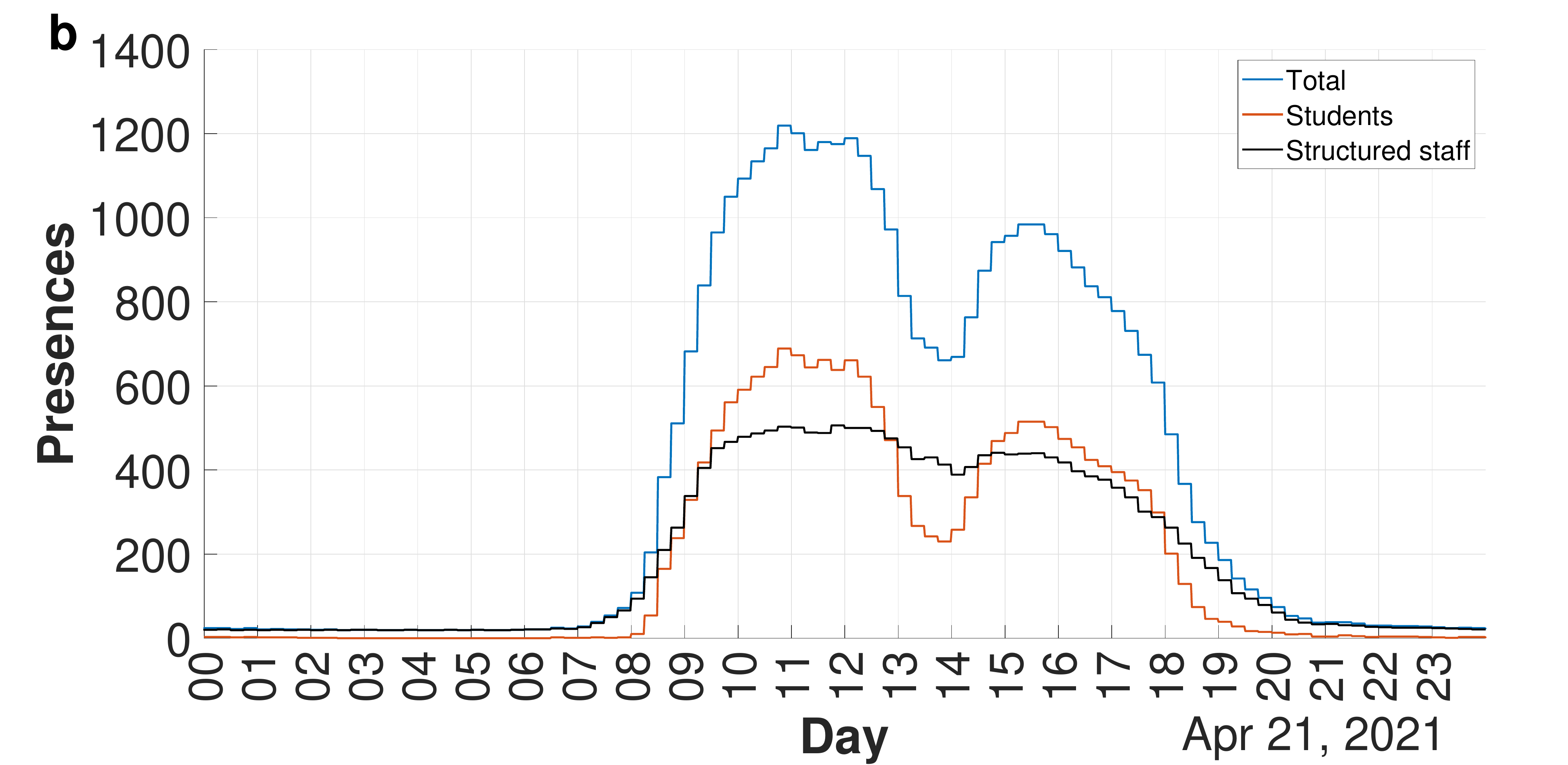}
\includegraphics[width=0.75\textwidth]{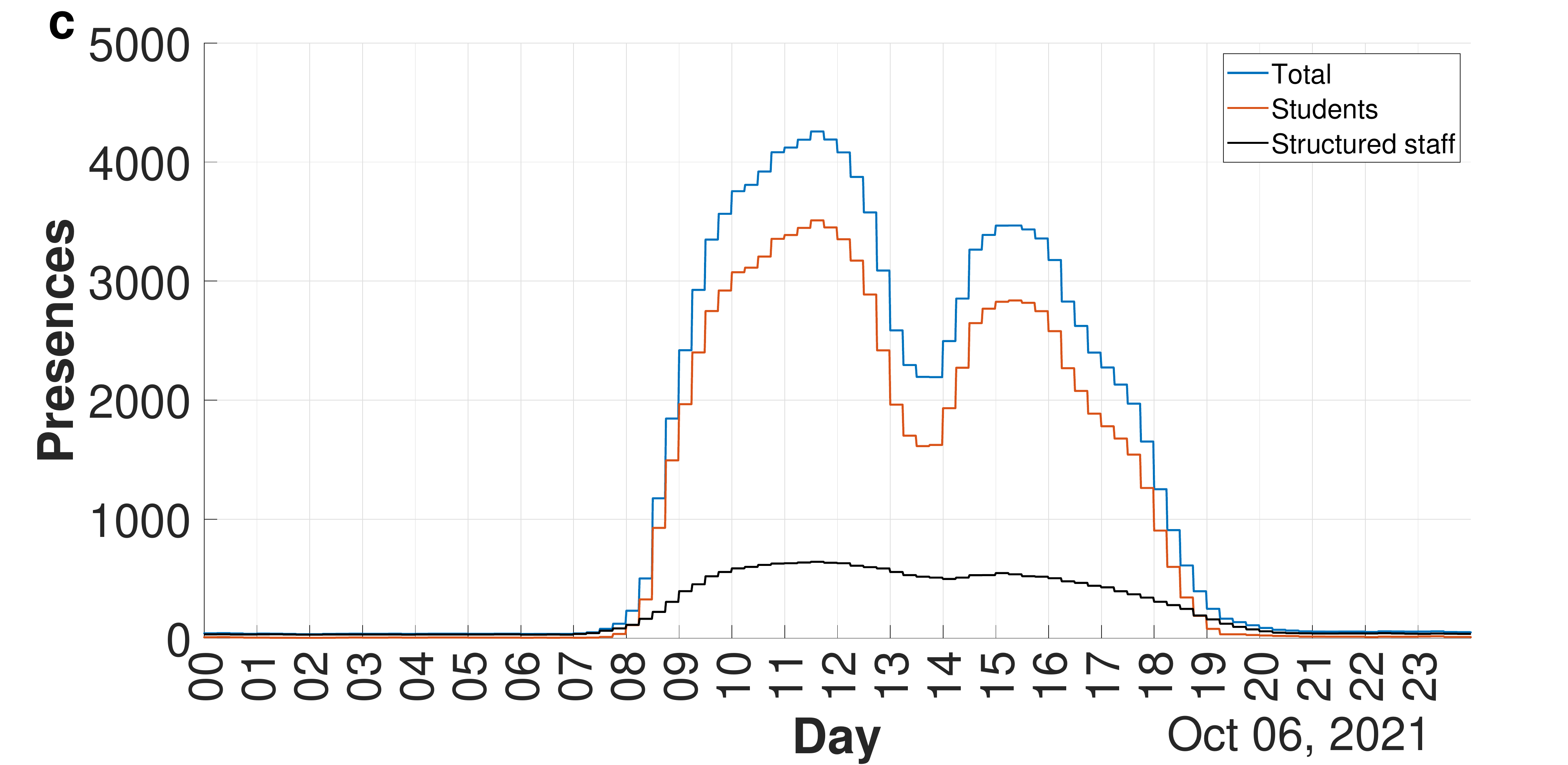}
\caption{{\bf Presences in Parma University during a typical working day.} 
     {\bf a} Time evolution of attendance during a typical working day in the closing regime. In this phase, staff members provides the larger contribution and the presences of students is very limited. In panel {\bf b} we see that students and structured staff provide a similar contribution, while in panel {\bf c} we notice that in the total opening phase the dynamical behavior is completely driven by student population.
     %From line red (students presences) we observe two different values of attendance for morning and afternoon: this is probably due to more lessons in the morning. For structured staff (black line), on the other hand, there are no different behavior between morning and afternoon. In panel {\bf b} we plot the presence in Campus during closing period (14th April 2021): the presence of students is dramatically decrease and we do not observe different behavior between morning and afternoon. Structured staff attendance plot is comparable with the same plot during opening phase.
     }
    \label{fig:figure2}
\end{figure*}

Fig ~\ref{fig:figure2} shows the attendance data in the University in a typical working day during the closing (panel {\bf a})
the partial opening (panel {\bf b}) and the total opening (panel {\bf c}) phases. In all cases, the curve rapidly increases in the morning, partially decreases during lunchtime, increases again in afternoon and decreases in the evening. We notice that in the closing phase staff members provide the largest contribution, which is similar in the partial opening phase. On the other hand, in the total opening phase the behavior is totally driven by the student population. Moreover the student population displays an oscillating behavior, with peaks corresponding to the morning and afternoon lessons, while the presence of staff members is distributed during the whole working hours.

\begin{figure*}[]
\centering
\includegraphics[width=0.75\textwidth]{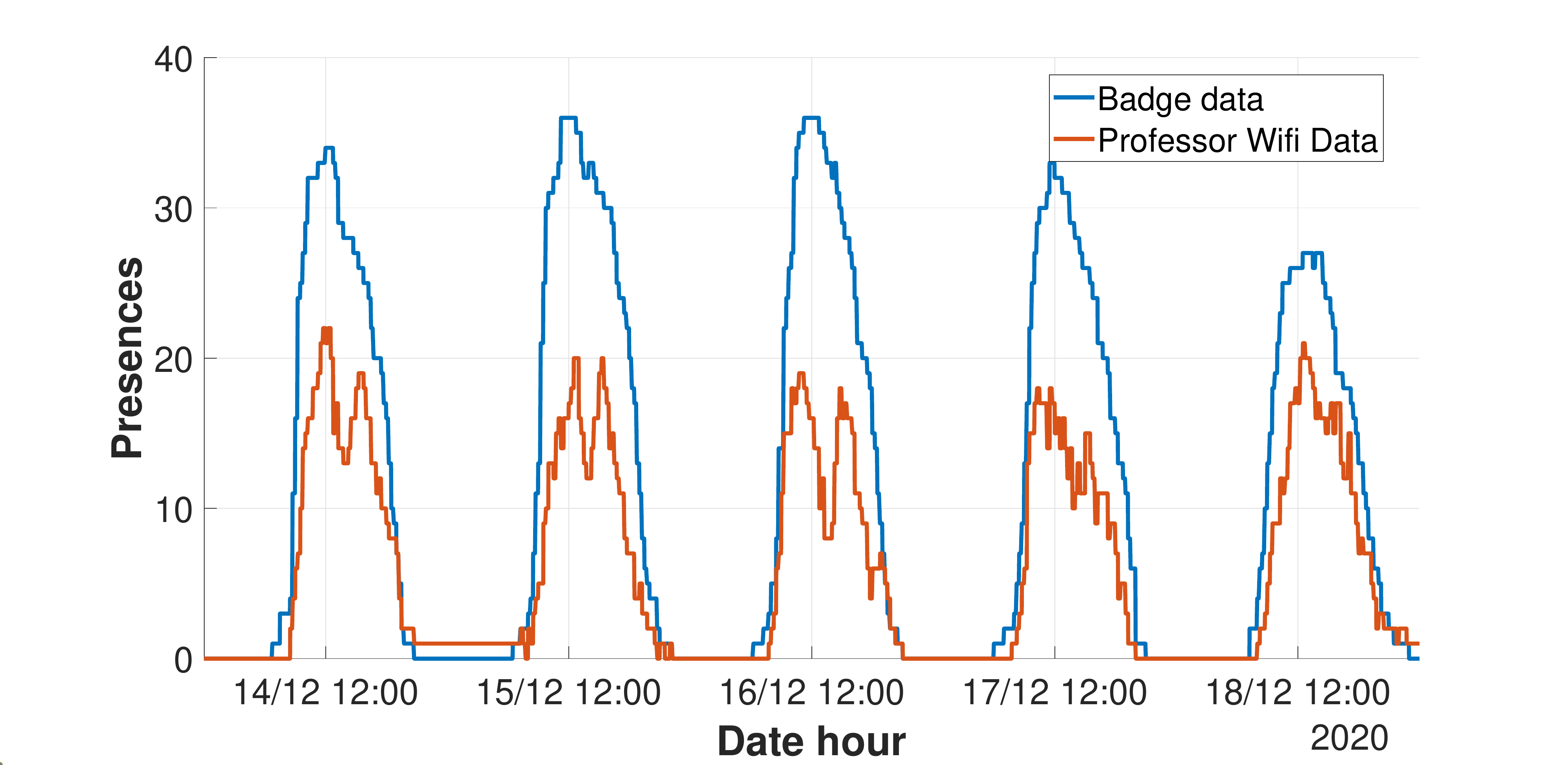}
\caption{{\bf Wi-Fi data calibration.} 
Comparison of the temporal evolution of attendance obtained from badge reader (blue line) data and from the WiFi data fro structured staff (red line) during a typical working week. The attendance obtained from WiFi is underestimated by about a factor $2$ compared to attendance provided by badge data.}
\label{fig:figure3}
\end{figure*}

As mentioned above, WiFi data only refer to individuals connected to the WiFi and this naturally leads to an underestimation of the total number of presences. To measure such an effect, we compare the WiFi data with data obtained from the badge access to a specific building (the Physics building). For this calibration we restrict our measure to structured staff in the partial opening phase. Indeed only in this case the use of the badge was compulsory. Fig~\ref{fig:figure3} shows that the attendance given by WiFi data are underestimates about by a factor $2$ compared to the attendance given by badges data. To obtain an analogous calibration for the student dataset, we compare WiFi data from APs near a specific classroom (``Aula Newton'') inside the Physics building with the online seat reservation needed to take part in the face-to-face lessons during the partial opening phase. Also in this case, we find an underestimation of the WiFi data by about a factor $2$ compared to online seat reservations.

\subsection{Data Limitations}

Passive WiFi data at the AP level is a coarse measure for people localization and co-location. Depending on multiple factors like signal strength and load on individual APs, devices in different rooms can connect to the same AP, or vice versa devices in the same room can connect to different APs \cite{Sure17}. A number of technologies exist to improve this kind of localization (see Section \ref{sec:related}) however they either require special sensing applications installed on the devices, or special sensors on the APs. None of them is widely available and deployed. Instead our goal is to investigate results coming from the WiFi-infrastructure ``AS-IS'' deployed nowadays in the majority building and public spaces.

Due to localization errors, some of the results described in the next section can present miscounts (e.g. people connected to the AP registered in a room that were actually another one). However the university buildings are large with large classrooms and halls \cite{Zhou16,4146826,Chen04} and over long observation periods these errors will likely average out, as it is more likely to be connect to the AP in the room where the device is actually present. Therefore we believe such mis-locations are limited in our scenario.
In this perspective, comparison between Wi-Fi data and badge readings reported in Figure \ref{fig:figure3} confirms that our method provides reliable results.

Moreover, we stress that our results are typically based on the comparison between different time periods characterized by different opening and closure phases. Therefore it is likely that possible errors are cancel-out, since the number of miscounts due to limitations of our technologies should affect the different phases in the same manner. 
Accordingly, we are confident that our results and conclusions hold despite data limitations, and that the advantages of using "standard" passive WiFi data, both in terms of widespread applicability and in terms of allowing long-term -- indefinite -- observation period outweighs the limitations. 

\subsection{Monitoring crowding with different approaches}
\label{sec:related}

There are several ways to use WiFi networks to localize and estimate people presence in an environment. We briefly summarize them and discuss applications of WiFi localization, and also other technologies, to analyze COVID-related crowding.

\subsubsection{Improved WiFi-based Measures for People Counting}

A number of technologies have been proposed to use the WiFi network as a mean to localize and estimate people presence. One of the key advantages of our proposal is to rely on the available WiFi-infrastructure ``AS-IS''. Even if the precision obtained in people localization is limited, the approach can be immediately applied to a wide range of environments, for a prolonged -- indefinite -- period of observation, providing sufficiently accurate measures of occupancy. Nevertheless, we think it is valuable to briefly survey existing approaches that could be used to improve WiFi localization with special purpose equipment that could be installed in critical areas. 

WiFi Real-Time Location System (RTLS) \citeauthor{cisco,9393471,7447750} uses multiple network signals (e.g., Received Signal Strength -- RSS, Angle of Arrival -- AoA)  from multiple neighboring APs to provide high precise localization of individual devices. Reports indicate that RTLS can achieve localization accuracy of less than 5-10m.

Even more advanced, the analysis of WiFi Channel State Information can detect and count people presence (actual people, not devices) in an environment. In particular \cite{9473673} analyzed this techniques (that requires specifically placed hardware) for COVID-Safe Occupancy Monitoring obtaining optimal performances.

A number of works, e.g., \cite{app1}, use specific apps to track and localize individuals based on WiFi signals. Continuous tracking via specific apps can notably improve localization accuracy, but they require users to actively install the application. Also approaches like the one in \cite{Sure17} would notably improve WiFi localization accuracy, but critically require WiFi signal sensing by the smartphone- therefore reacquiring specific (monitoring) apps to be installed.  Similarly, \cite{8936381} applies advanced estimation mechanisms to calibrate WiFi measures on the basis of camera footage. After a calibration phase, they achieve improved performance on people counting.

This kind of technology would notably improve our estimates, but it is not widespread in WiFi deployments, or it needs the involvement of multiple users, therefore it cannot be directly applied without notable investments.

\subsubsection{Systems for the Analysis of Distancing}

There are also some available approaches to detect distancing in general, not WiFi based, which we briefly review here.

Mobile phone applications have been one of the most used technologies to monitor social distancing and tracking COVID diffusion \cite{9144194,10.1145/3494529}. In this context Bluetooth inferred proximity \cite{Eagle06, Sure17} is one of the  mainly used technologies. These approaches basically rely on mobile phones localization and/or detection of nearby devices to detect social distancing. As well documented also in mainstream news, the main issue with this technology is about user acceptance and willingness to actively install such applications.

Another important body of work is about monitoring large user population via mobile phone network \cite{mobile}. A number of network signals are recorded by mobile phone operators and can be used to effectively localize individuals and analyze social distancing. While these applications can cover large areas, network-based localization accuracy is often too coarse to detect social distancing.

The recent advancements of computer vision technology and widespread deployment of fixed and mobile cameras (e.g., on drones) attracted a number of works analyzing social distancing on the basis of video analysis \cite{9138385}. From a technological point of view this is probably one of the best solution, but it is hindered by strong privacy  concerns about accessing multiple video feeds.  

More in general, \cite{10.1145/3514137} introduces several challenges in multiple technologies to contact tracing and analysis of social distances.

\subsubsection{WiFi-based Systems for the Analysis of Distancing}

Restricting our focus on WiFi-based approaches to monitor distancing, we find different directions.

The work in \cite{Zak20} analyzes user occupancy and mobility via deployed WiFi infrastructure to help institutions to monitor and maintain safety compliance according to the public health guidelines. While the overall motivations and goals of this work are very similar to ours, there are some important differences with regard to the metrics derived from WiFi data. In our analysis, we strove to be fully GDPR compliant. Therefore analysis/applications requiring maintaining user-id for a prolonged time - like the mobility analysis conducted in this related work - is actually unfeasible under GDPR. Vice versa, we identified novel, useful and GDPR-compliant analyses - e.g., the AP different links per day -  that can better support institutions. Similarly, \cite{Mu20,10.1145/3384419.3430598}  describes a similar analysis with the aforementioned differences with our approach.

Analogously, \cite{Gup20} presents a system to observe individuals and spaces to implement policies for social distancing and contact tracing using WiFi connectivity data in a passive and privacy-preserving manner.

One of the most interesting related work is 
\cite{Swa21}. They start from motivations similar to ours, and compute analogous co-location metrics. However, their approach does not take into consideration ``forced'' distancing in some environments and cannot differentiate between ``safe'' areas with lots of co-locations but distanced between each other (e.g., classes) and ``unsafe'' areas where distancing is not enforced (e.g., hallways). On the other hand, our work presents a compelling analysis of the localized closures policies that can be enacted on the basis of the obtained results. Such policies could be effectively implemented also over our data and approach.

Similarly, the recent work \cite{10.1145/3516524} discusses passive WiFi-monitoring in a campus scenario to monitor social distances. While goals, motivations and technologies are very similar to ours, our work focuses on a much longer monitoring period to analyze the impact of different policies enacted to contain COVID epidemics. Moreover, our work applies novel analysis to better measure epidemiology-measures -- see Section 4.2 and 4.3.

\section{Results}

\subsection{The simplex size distribution}

During the Covid-19 pandemic, great attention has been devoted into limiting large gatherings of people especially in situation where the tracing of the attendance is a difficult task.  
In this framework WiFi dataset provides a natural way of monitoring the presence of large groups (simplices) within a certain area by considering the number of people simultaneously connected to the same  AP. As a first step, we extract the group size $s$ distribution $P(s)$ from WiFi data connections during working hour (from 8.00 am to 7.00 pm). 

In social systems, the duration in time of face to face contacts typically displays a broad distribution with non trivial behaviors \cite{stehle10,zhao11}. However, the infection process occurs on a characteristic timescale, which for COVID-19 has been estimated to be of about 15 minutes, for the first variants \cite{ECDC_ct}. We therefore 
define a simplex of size $s$ a group of $s$ persons which are connected to the same AP for at least 15 minutes (the same time interval used in contact tracing apps). Since this choice can influence the cluster size distribution, we also verify that our main results are qualitatively independent of this choice, by varying the time interval from 5 to 30 minutes. In particular,
for each AP, we split working hours (from 8.00 am to 7.00 pm) into 15 minutes intervals and for each interval we find the number of users that were connected to that AP for the entire time intervals: this number corresponds to the simplex size.  Disconnections from a single AP shorter than 5 minutes have been discarded from our data set.
We distinguish the three phases of the pandemic period with different levels of restriction (closing, partially opening and total opening phases) and obtain three different distributions of group sizes.

\begin{figure*}[]
\centering
\includegraphics[width=.75\textwidth]{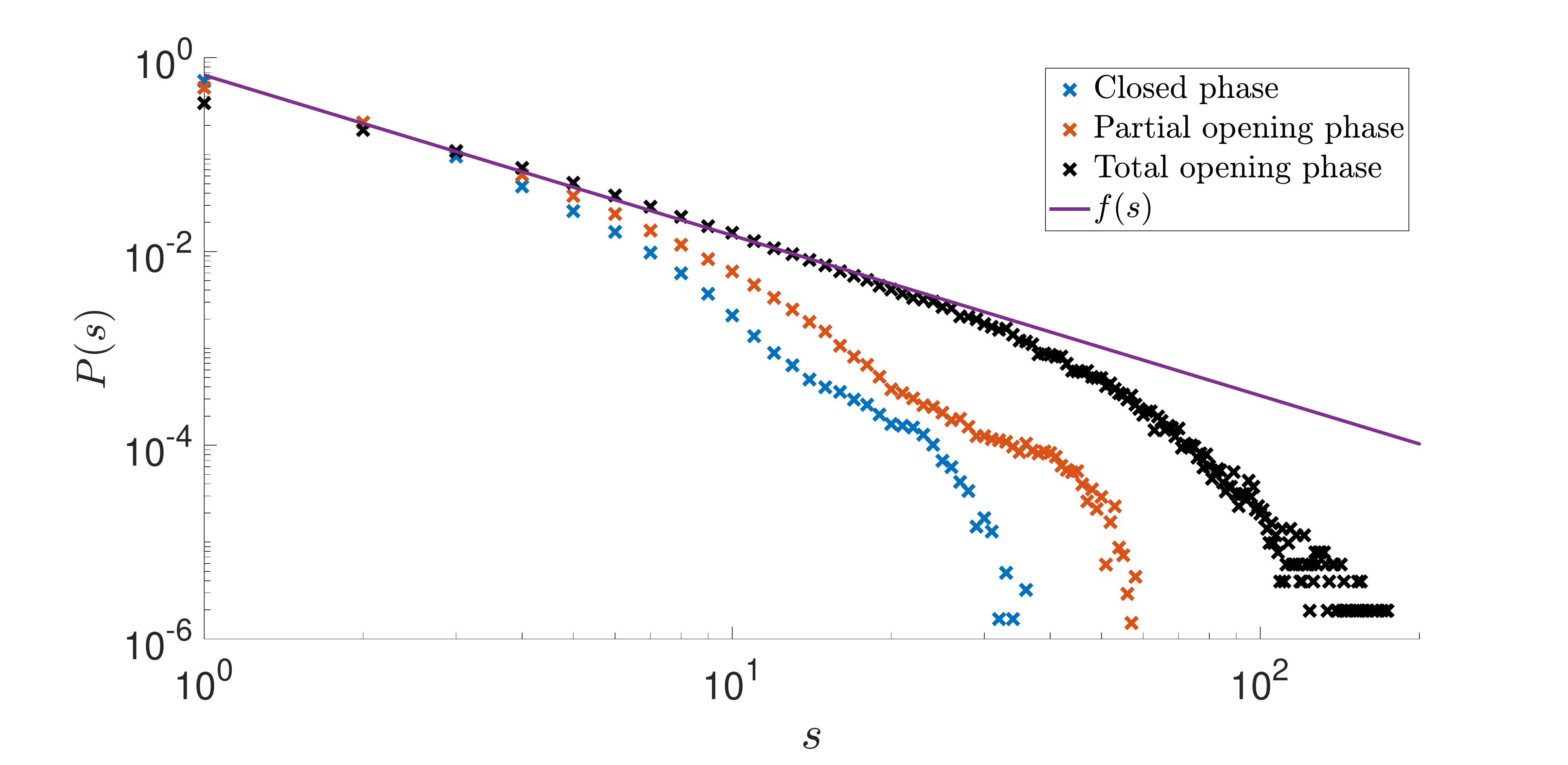}
\caption{{\bf University group size distribution.}
Probability $P(s)$ to find a group of $s$ people connected to the same AP for almost 15 consecutive minutes in log-log scale. We plot the group size probability distribution for each phase with different restriction regimes. The distribution $P(s)$ in the total opening phase is compatible with a power law $f(s) \propto s^{- \nu }$ with an exponent $\nu \approx 1.65$.
%when the students could only enter the University for laboratories, red line is the probability during partially opening phase with the partial return of lessons in partial presence. 
}
\label{fig:figure4}
\end{figure*}

\begin{figure*}[]
\centering
    \includegraphics[width=0.9\textwidth]{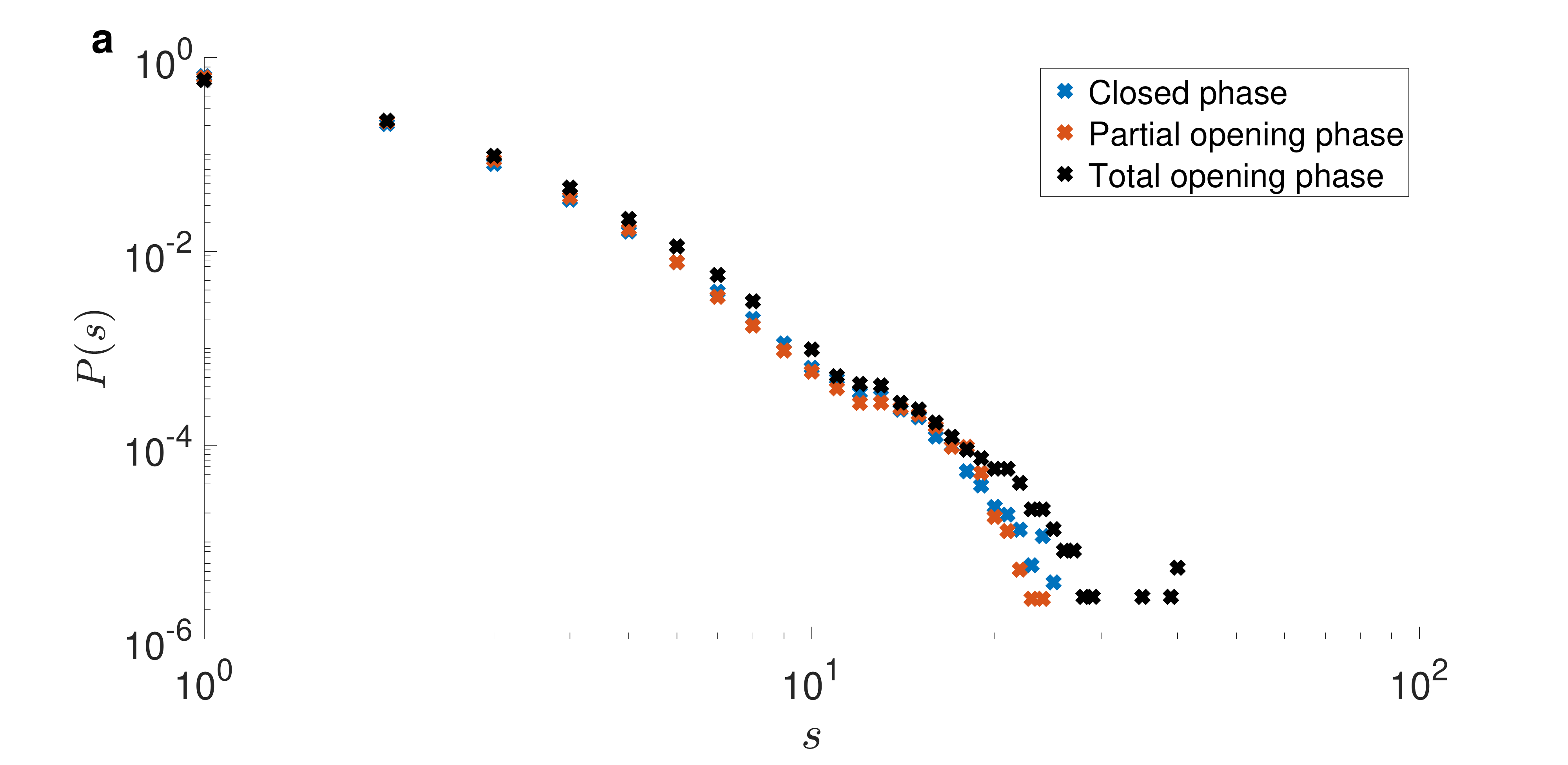}
    \includegraphics[width=0.9\textwidth]{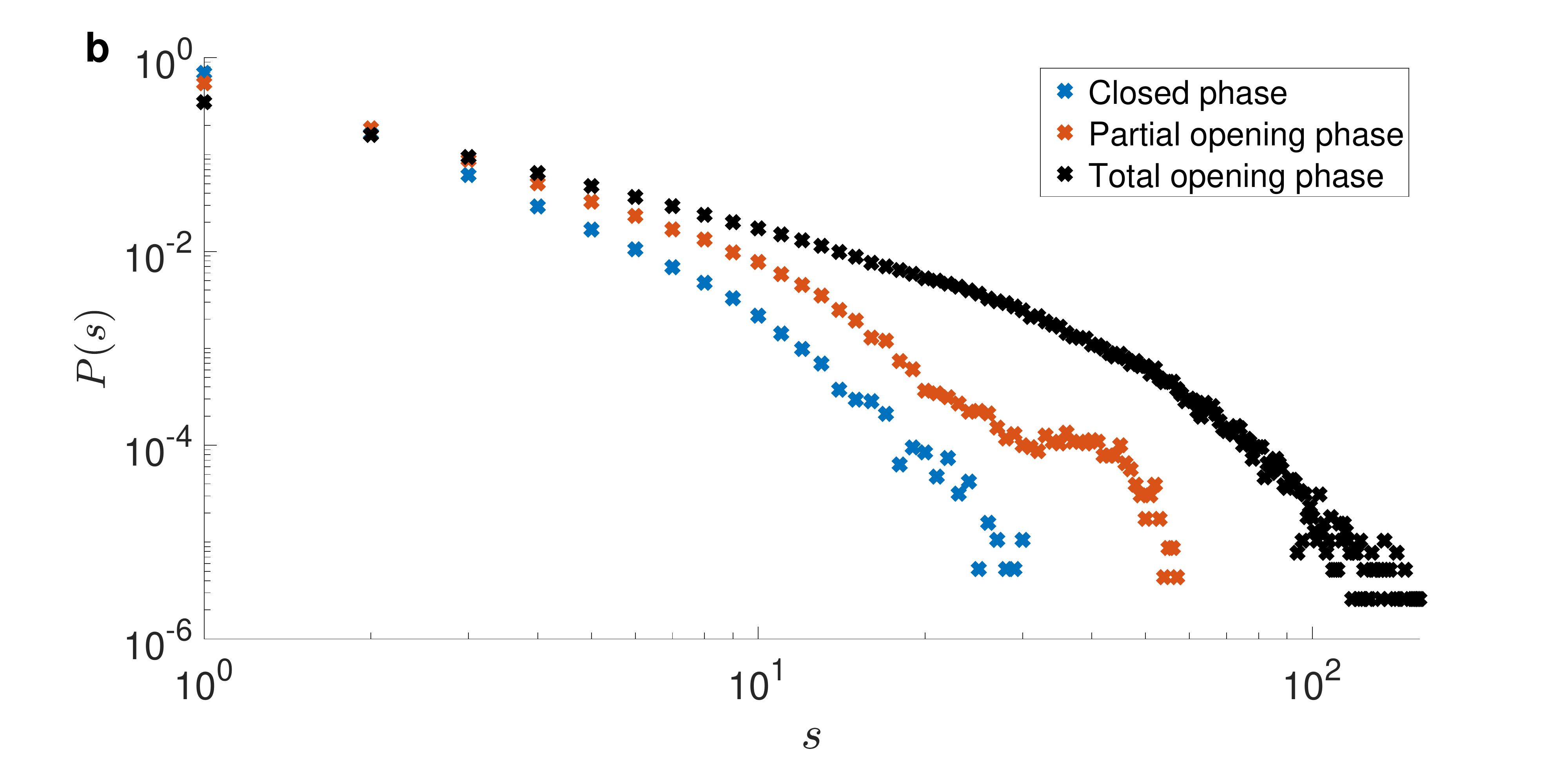}
\caption{{\bf Group size distributions comparison.} Panel {\bf a} show the contribution of university staff to the group size distributions in the three phases with different restriction regimes. The effect of the restrictions in the closing and partial opening phase are very limited. On the other hand, we note in panel {\bf b} that the restrictions have a strong impact on the behavior of students population, whose group size distribution is strongly modified in the different phases.}
\label{fig:figure5}
\end{figure*}

Fig~\ref{fig:figure4} clearly shows that a broader distribution $P(s)$ is observed in the opening phases, and the maximum group size grows from $s_{max}^C=36$, to  $s_{max}^{PO}=58$ and to $s_{max}^{TO}=174$ in the closing, partial opening and opening phases, respectively. In particular, the distribution $P(s)$ in the total opening phase is compatible with a power law $P(s) \propto s^{-\nu}$ with an exponent $\nu \approx 1.65$. In
Fig~\ref{fig:figure5} we plot the different contribution of University staff and students to the group size distribution. Again, the plot clarifies that the restrictions in the closing and partial opening period mainly affect the behavior of the student population, whose simplex distribution is strongly modified in the different phases, while for staff members the differences are very limited. 

\subsection{Epidemics on simplicial temporal networks}
\label{sub:R0}

%Epidemic spreading within a population can be described adopting standard compartmental models \cite{pastor2015}. In the Susceptible-Infected-Recovered (SIR) model, a susceptible individual \emph{S}, is infected through a contact with any infected \emph{I} agent with probability $\lambda$:
%$I+S \rightarrow I+I$. 
%Infected \emph{I} individuals recovers at rate $\mu$,
%$I \rightarrow R$, 
%and recovered individuals $R$ are immune. The epidemic propagation is governed by the basic reproduction number $R_0$, which depends on the parameters of the model and also on the structure of the interactions among the agents: if $R_0<1$ the epidemic does not spread and only a finite number of people is infected, while for $R_0>1$ the epidemics displays an exponentially growing outbreak, eventually infecting a finite fraction of the whole population.
Epidemic compartmental models have recently been extended to time-evolving networks driven by the activity of nodes, in the framework of activity-driven networks \cite{Perra2012,tizzani2018,Mancastroppa2019}. In particular, in simplicial activity driven networks \cite{petri2018simplicial,mancastroppa2021sideward} nodes are grouped into fully connected clusters of different sizes, which are continuously activated and destroyed. In this context, parameters tailored on SARS-CoV-2 transmission \cite{michael2021,Guan2020,who_china_report} have been considered. 

In the Susceptible-Infected-Recovered (SIR) model on activity driven simplicial networks \cite{petri2018simplicial,mancastroppa2021sideward}, the interaction network evolves by activating simplices at rate $a$, the simplex activity; when a simplex (clique) of size $s$ is active, $s$ nodes are chosen uniformly at random to participate in the simplex, producing
$s(s-1)/2$ interactions. Then the cluster is destroyed and the process is iterated.
The simplex size $s$ is drawn from the distribution $P(s)$ of the sizes of the clusters, which models the heterogeneity in the size of gatherings. Each susceptible node of the cluster is infected with probability $\lambda$ by the infected nodes $I$ belonging to the same cluster. Then each nodes randomly recovers at rate $\mu$, as in the original SIR. The epidemic propagation is governed by the basic reproduction number $R_0$: if $R_0<1$ the epidemic does not spread and only a finite number of people is infected, while for $R_0>1$ the epidemics displays an exponentially growing outbreak, eventually infecting a finite fraction of the whole population.
Since in the network evolution each simplex is randomly reconstructed at each activation so that the connections among the agents are continuously reshuffled, a mean field approach exactly describes the evolution of the system: 
\begin{eqnarray}
\partial_t S(t) & = & - S(t) \int  a s P(s)  \left[1-\left(1-\lambda I(s) \right)^{s-1}\right] ds\nonumber \\
\partial_t I(t) &=& -\mu I(t) + S(t) \int  a s P(s) \left[1-\left(1-\lambda I(s)
\right)^{s-1}\right] ds  \label{eq:MF}\\
\partial_t R(t) &=& \mu I(t) \nonumber
\end{eqnarray}
where $a$ is the activation rate of a simplex (i.e. the number of simplices in the system per unit of time), $\lambda$ is the probability that an active link transmits the disease, and $\mu$ is the recovery rate of a node. $S(t)$, $I(t)$ and $R(t)$ represent the probabilities for a node to be susceptible, infected or recovered at time $t$; the normalization condition implies that $S(t)+I(t)+R(t)=1$. In Eq.s \ref{eq:MF}, the term $\mu I(t)$ represents the recovery rate of the process $I\rightarrow R$ and the integral describes the infection process: i.e. $a s S(t)P(s)$ is the activation rates of susceptible nodes into a cluster of size $s$ and $\left[1-\left(1-\lambda I(s)
\right)^{s-1}\right]$ is the probability that such activated susceptible node is infected by one of 
the remaining  $s-1$ individuals of the cluster; finally we sum (integrate) over all the possible cluster sizes.
The stability of the solution where all nodes are susceptible (i.e. $S=1$, $I=0$ and $R=0$) can be studied by linearization. In particular, the linearization of the second of Eq.s \ref{eq:MF} gives 
\begin{equation}
\partial_t I(t) = (-\mu + a \lambda \langle s(s-1)\rangle)  I(t).
\label{eq:MF_lin}
\end{equation}
with $\langle s(s-1)\rangle=\int s(s-1) P(s) ds $ and the solution with $I=0$, $R=0$ and $S=1$ is stable only if 
$\mu > a \lambda \langle s(s-1)\rangle$ and the basic reproduction number reads: 
\begin{equation}
R_0=\frac{a \lambda \langle s(s-1)\rangle}{\mu}.
    \label{eqn:threshold}
\end{equation}

The measure of the parameters in Eq. \ref{eqn:threshold} is a difficult task, however the expression for $R_O$ can be used to obtain a first estimate on the effect of a variation of the network connectivity on the epidemic propagation.

Since $s (s - 1)/2$ is the number of links in a fully connected simplex of size $s$, Eq.\ref{eqn:threshold} states that $R_0$ is proportional to the number of connections present in the system. As in the activity driven models, individuals of a cluster are randomly reshuffled at each time steps, correlations and memory effect characterizing the behavior of real temporal networks are not included in this approach \cite{tizzani2018,Ubaldi2016}.

Eq. \ref{eqn:threshold}  allows for a comparison among the three phases to estimate the change in the basic reproduction number due to the different size distribution $P(s)$ in the closing, partial opening and total opening phases (Fig. \ref{fig:figure4}). 
In particular,  $R_0$ increases from the closing to the partial opening phase as:
\begin{equation}
    \frac{R_{0,PO}}{R_{0,C}} = \frac{\left<s(s - 1)\right>_{PO}}{\left<s(s - 1)\right>_{C}} \approx 2.63
    \label{eqn:ratio}
\end{equation}
while going from the partial opening to the total opening phase implies:
\begin{equation}
    \frac{R_{0,TO}}{R_{0,PO}} = \frac{\left<s(s - 1)\right>_{TO}}{\left<s(s - 1)\right>_{PO}} \approx 13.03
    \label{eqn:thresholdThirdPhase}
\end{equation}

We verify that our estimates for the ratio ratios in Eq.s (\ref{eqn:ratio},\ref{eqn:thresholdThirdPhase}) are robust if we consider different time intervals to define the simplices from  our data-set. In particular, for a time interval of 5 minutes we obtain: 2.91 and 12.36 while for a time interval of 30 minutes 2.78 and 12.47 for the ratios in Eq. \eqref{eqn:ratio} and \eqref{eqn:thresholdThirdPhase} respectively.

We notice that the ratio between the total and partial opening phases is much larger than the ratio between the partial and closing phase. Therefore, while one could imagine to control the spreading with tracing measures in the transition from the closing to  the  partial  opening  period   \cite{mancastroppa2021sideward},
the same kind of control would turn out to be ineffective in the transition to the total opening. However, we point out that in the total opening phase a large percentage of the population had been vaccinated (about $80\%$ in Italy) and the use of the Green-Pass (mandatory for the access to the University buildings) implies that only vaccinated or tested people are admitted to the University buildings. Our data show that these additional measures are clearly necessary to limit the propagation in the total opening phase, due to the huge increase in the potentially dangerous contacts. Moreover, we notice that beside tracing procedures, use of masks and social distancing are still active even in the total opening period.

\subsection{The link distribution at the Access Points}
So far we analyzed the sizes of groups that form in the University and we compared the distributions in the three phases with different restrictions. In particular the data of simplex sizes can be also used to identify the most critical areas in the University, where large gatherings more frequent. 
However, the formation of large groups is not the only relevant information to determine if an AP is critical. Indeed, if the large gatherings are due e.g. to face-to-face lessons, we expect that the groups are stable for the whole duration of the lesson and that the contagion could be effectively traced (e.g. in the partially open period,  an online seats reservation procedure was active). On the other hand, there could be places (such as an atrium) where small groups form but they are continuously reshuffled. These places are typically dangerous because they host a high number of different contacts which are very difficult to trace. For these reasons we introduce a different characterization of the APs in order to find places where a large variety of contacts may occurs.
In particular we define the {\emph{daily links}} $l_i$ of an AP as the number of contacts per day formed for more than 15 consecutive minutes between two different users in the same location. The index $i$ hereafter labels the different APs. In this framework, if two users meet two or more times at the same point but at a different time, this is counted only once. Again, in the counting of users pairs we have considered only the working time.

%\begin{figure*}
%\centering
%\includegraphics[width = 0.8 \textwidth]{figure_pdf/campus_activity.pdf}
%    \caption{\csentence{Parma University Campus links distribution.} Sum of all AP daily links inside Parma University Campus in log-log scale plot and exponential bin width. Blue points represent the activity during closing period while black points represent the opening period. During partial opening we see greater activity than closing phase.}
%\label{fig:figure6}
%\end{figure*}

\begin{figure*}[]
\centering
\includegraphics[width=70mm]{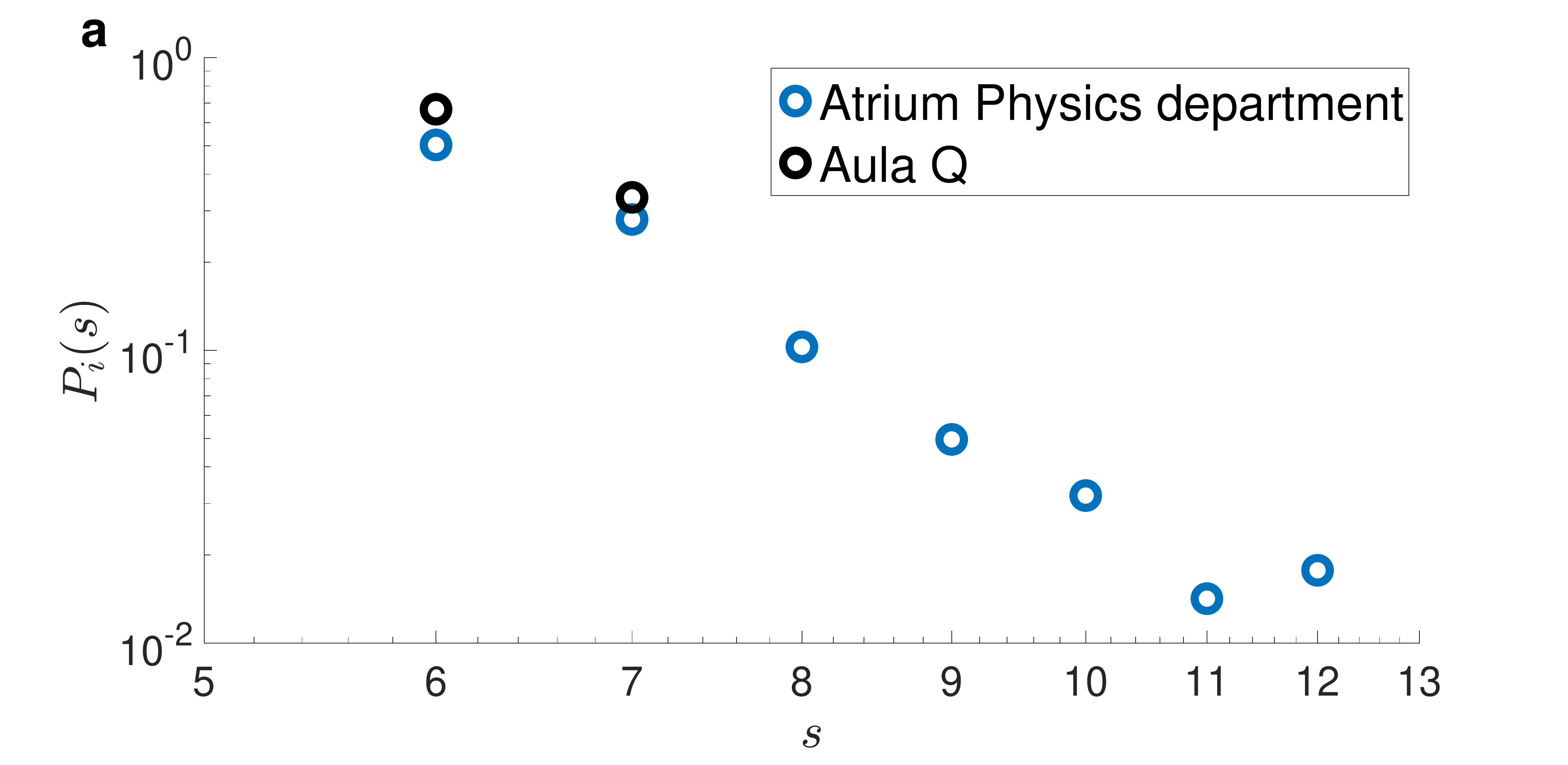}
    \includegraphics[width=70mm]{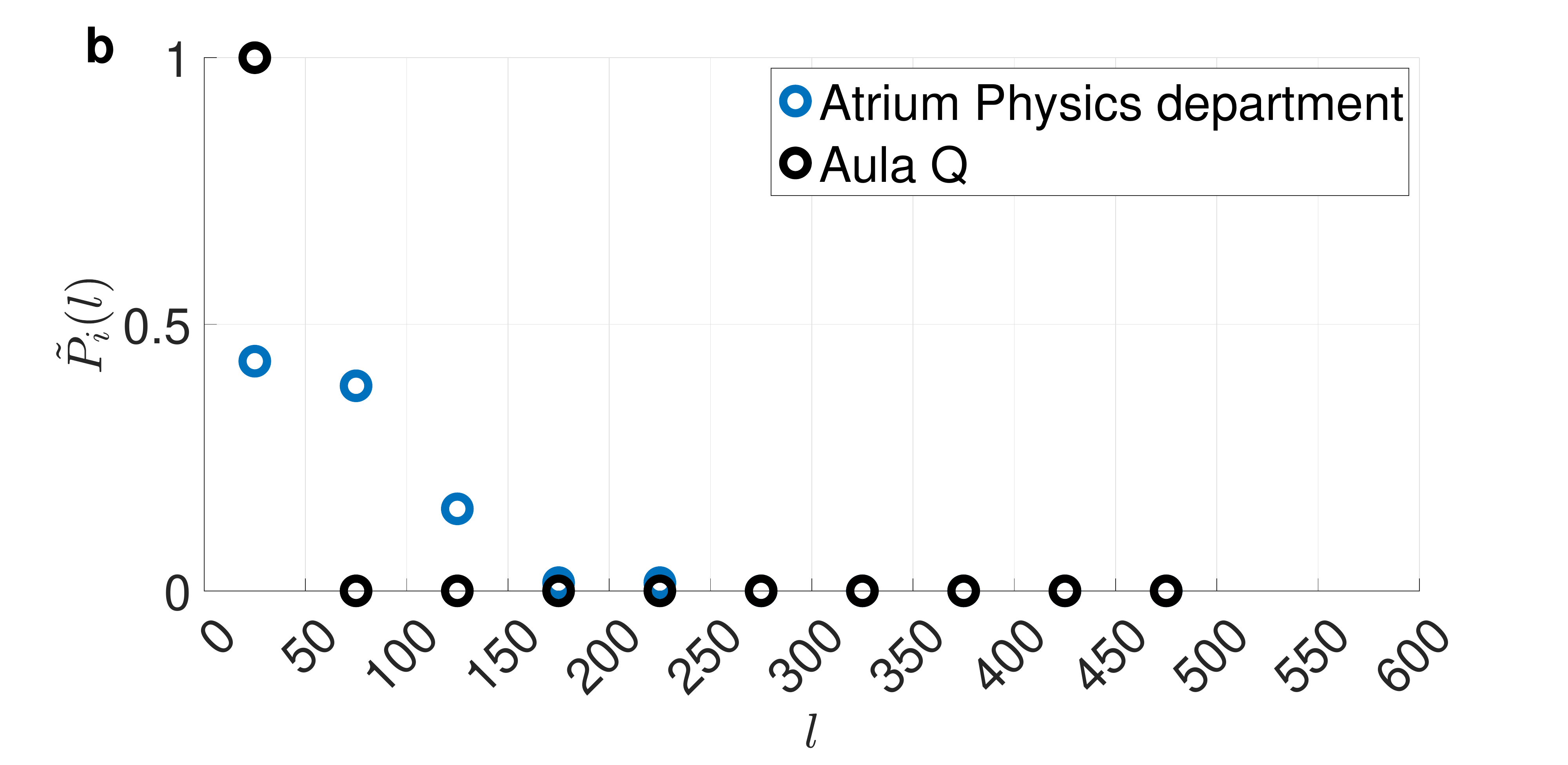}
    \includegraphics[width=70mm]{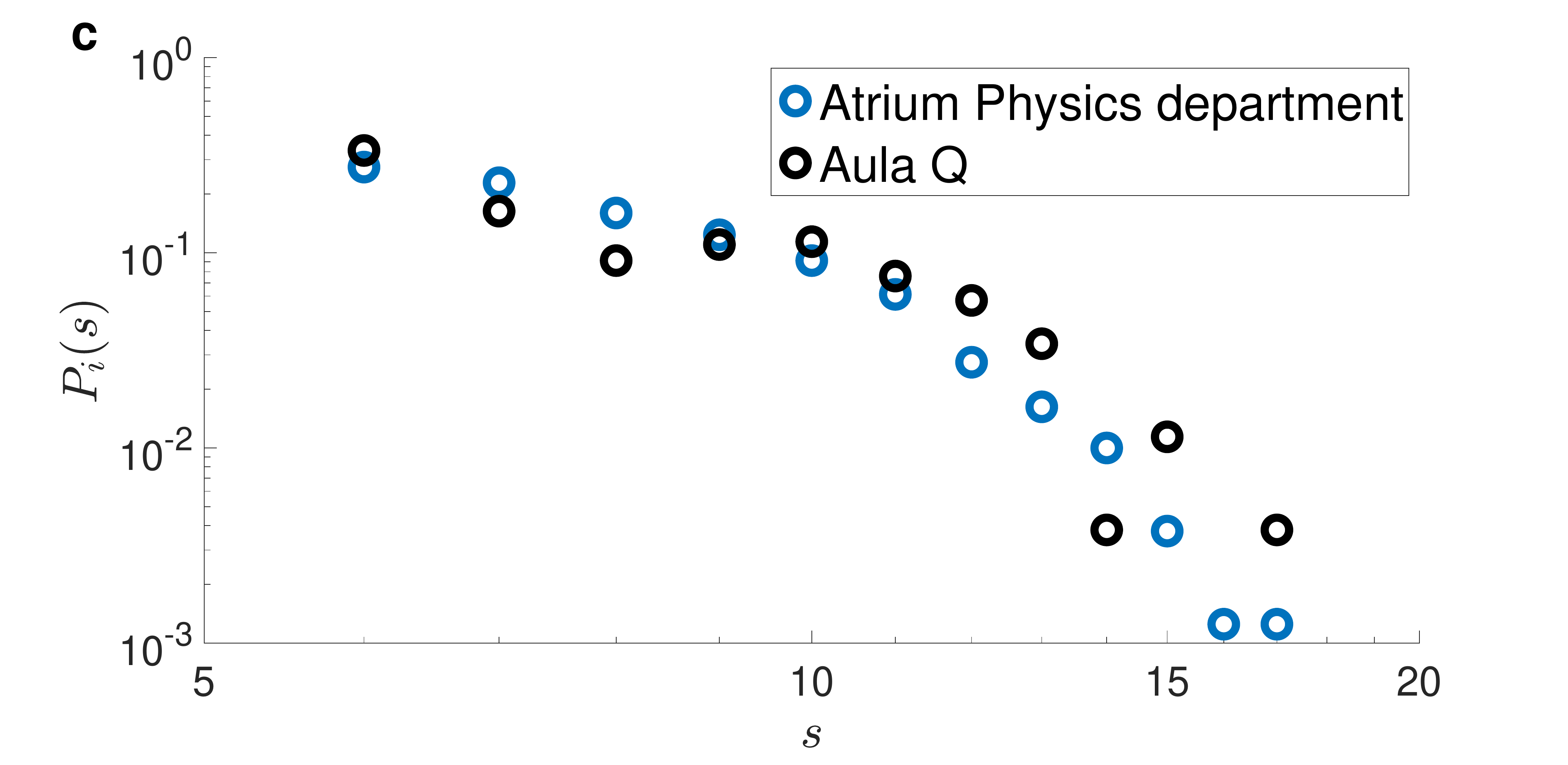}
    \includegraphics[width=70mm]{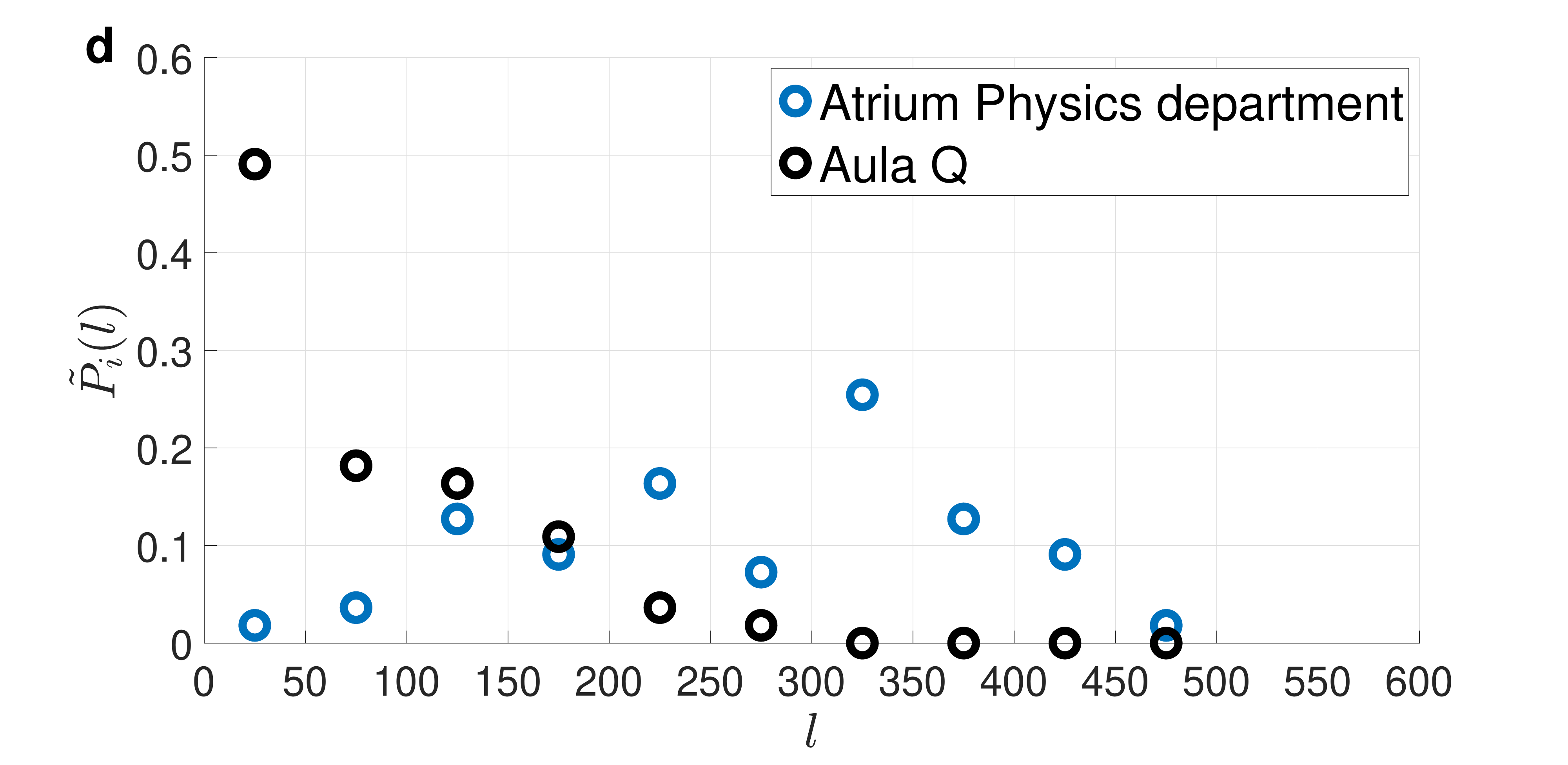}
    \includegraphics[width=70mm]{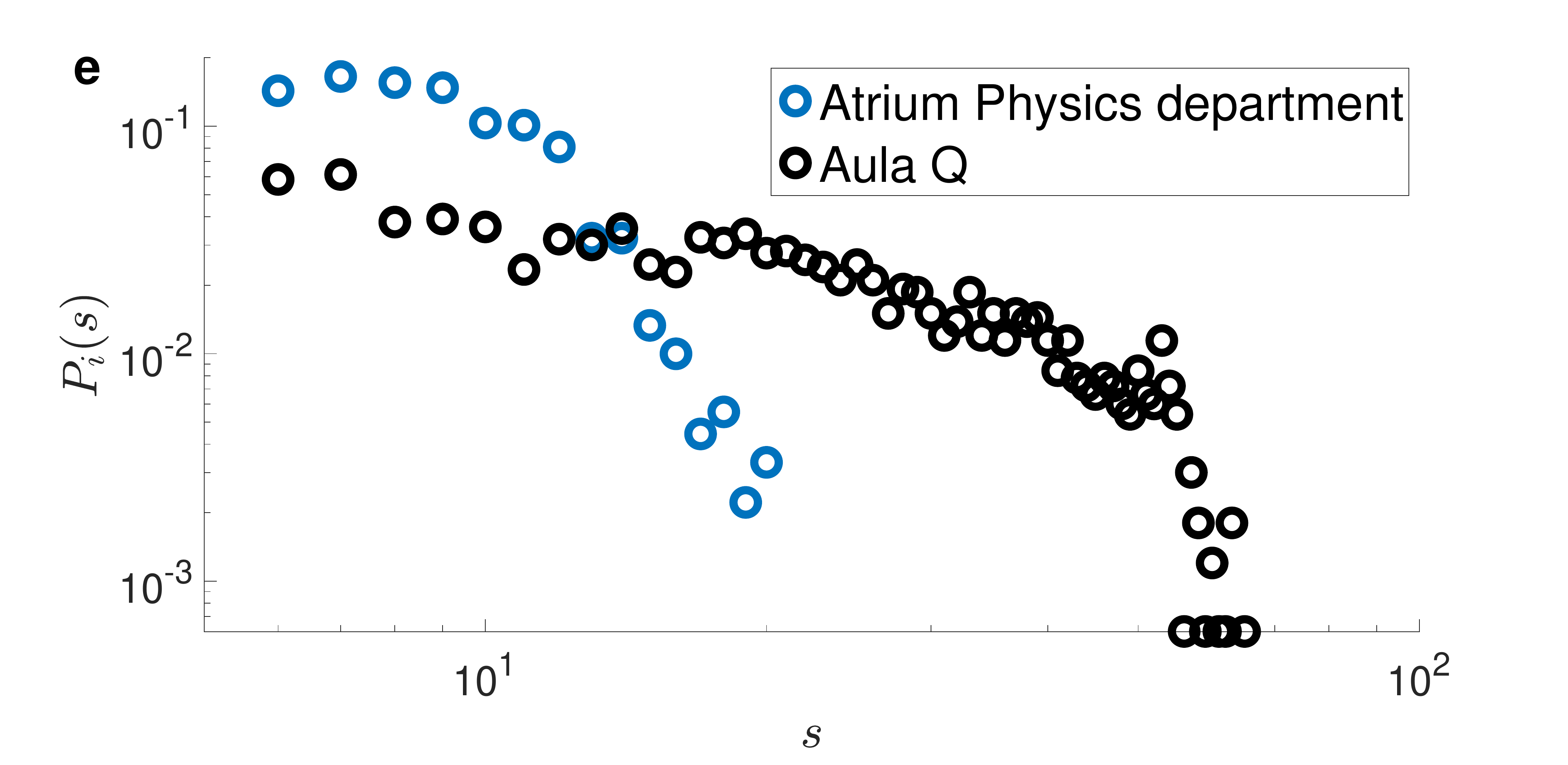}
    \includegraphics[width=70mm]{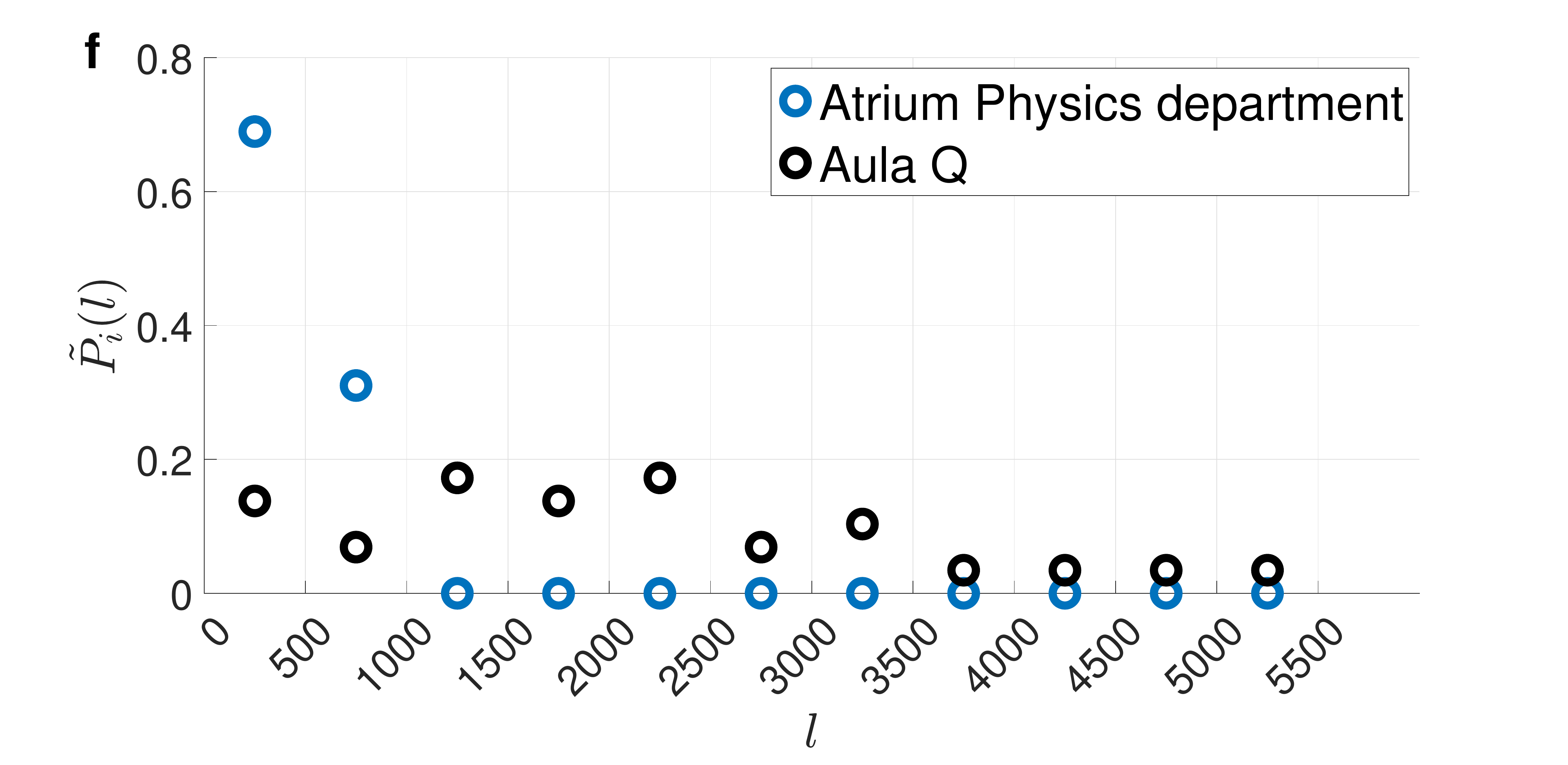}
    \includegraphics[width=70mm]{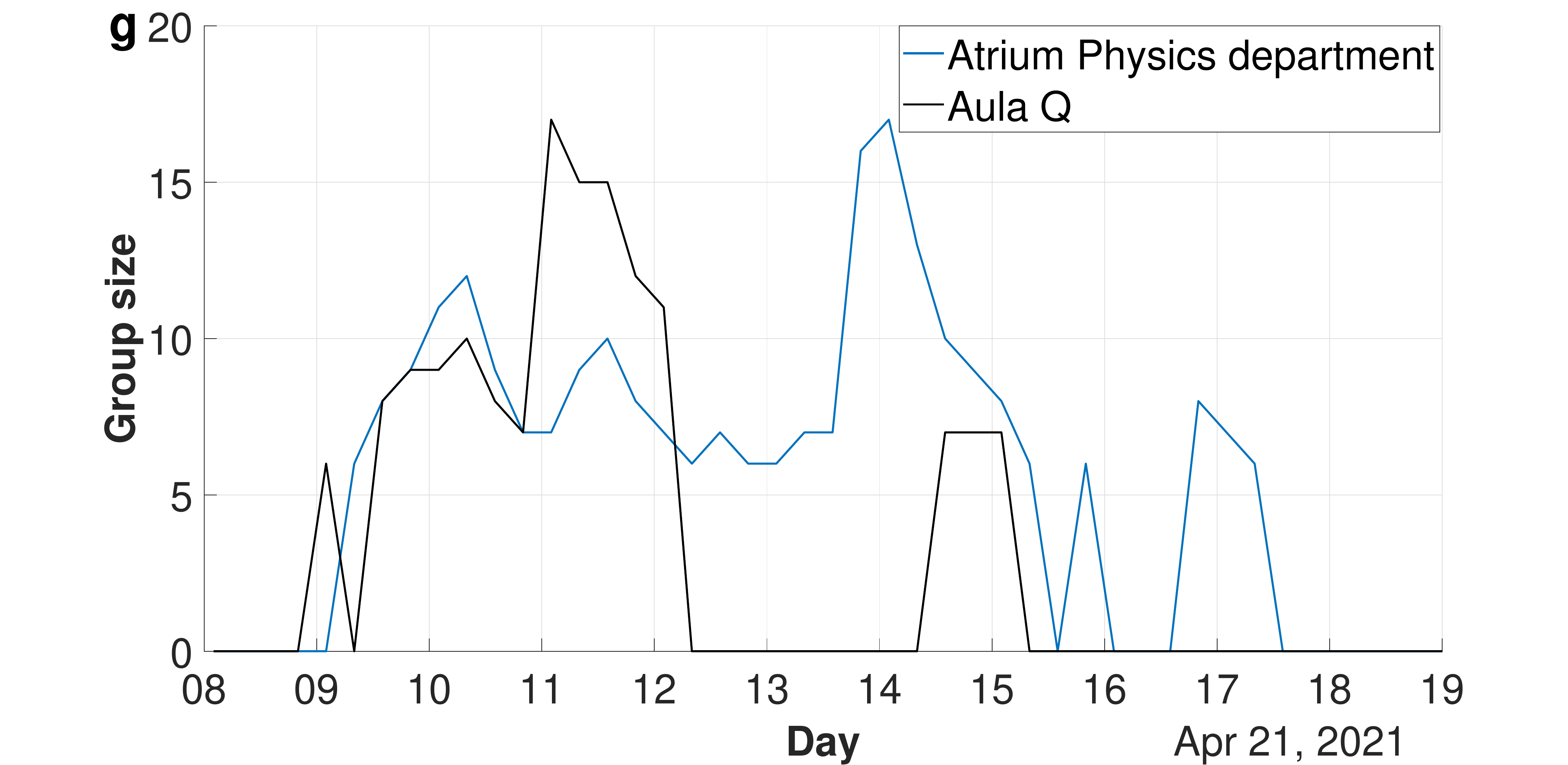}
    \includegraphics[width=70mm]{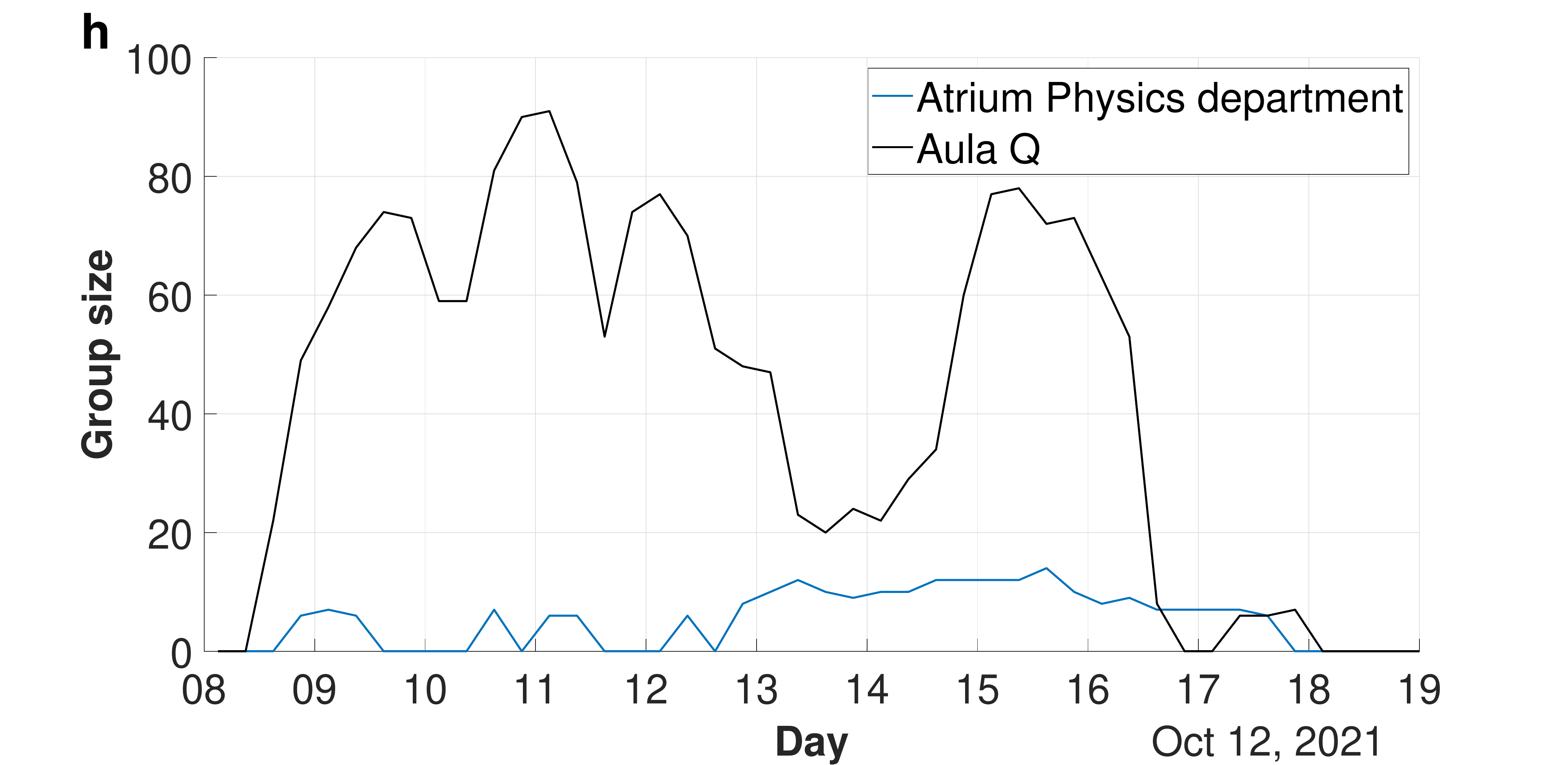}
\caption{{\bf Comparison between different APs of the simplex size and daily links measures.} We consider two APs installed in two different types of areas; one placed in classroom and the other in the atrium of Physics Department, and we compare their group size and daily link distributions. {\bf a} and {\bf b}: Plot of group size distribution and daily link during the closing phase. The AP placed in classroom shows that small groups are formed due to restrictions. {\bf c} and {\bf d}: For the same APs, we plot the distributions obtained during the partial opening phase. The two APs have a similar group size distribution but different daily link distributions. This shows that the groups in the classroom are more stable than in the atrium, where the groups are continuously reshuffled. {\bf e} and {\bf f}: The plots show that large gatherings are formed in the classroom due to the return of face-to-face lessons.  {\bf g} and {\bf h}: Time evolution of the group size for the two APs during typical working day in partial opening and total opening respectively. For small groups with size $s < 6$ we have set the group size to $0$. }
%\caption{\csentence{Comparison between different APs of the simplex size and daily links measures.} Comparison two specific APs; one placed in classroom and the other in the atrium of Physics Department. Panels {\bf a}, {\bf c} and {\bf e} show the group size distributions in the closing, partial opening and opening phases respectively. The plots show that two APs have a similar group size distribution during the closing and partial opening phases. The return of face-to-face lessons causes differences between the distributions of the two APs. Panels {\bf b}, {\bf d} and {\bf f} show that the daily link distributions are very different in all phases, highlighting a different behavioral dynamics of the users in the two areas where the two APs have been installed.}
\label{fig:figure7}
\end{figure*}

\begin{figure*}[]
\centering
\includegraphics[width=.75\textwidth]{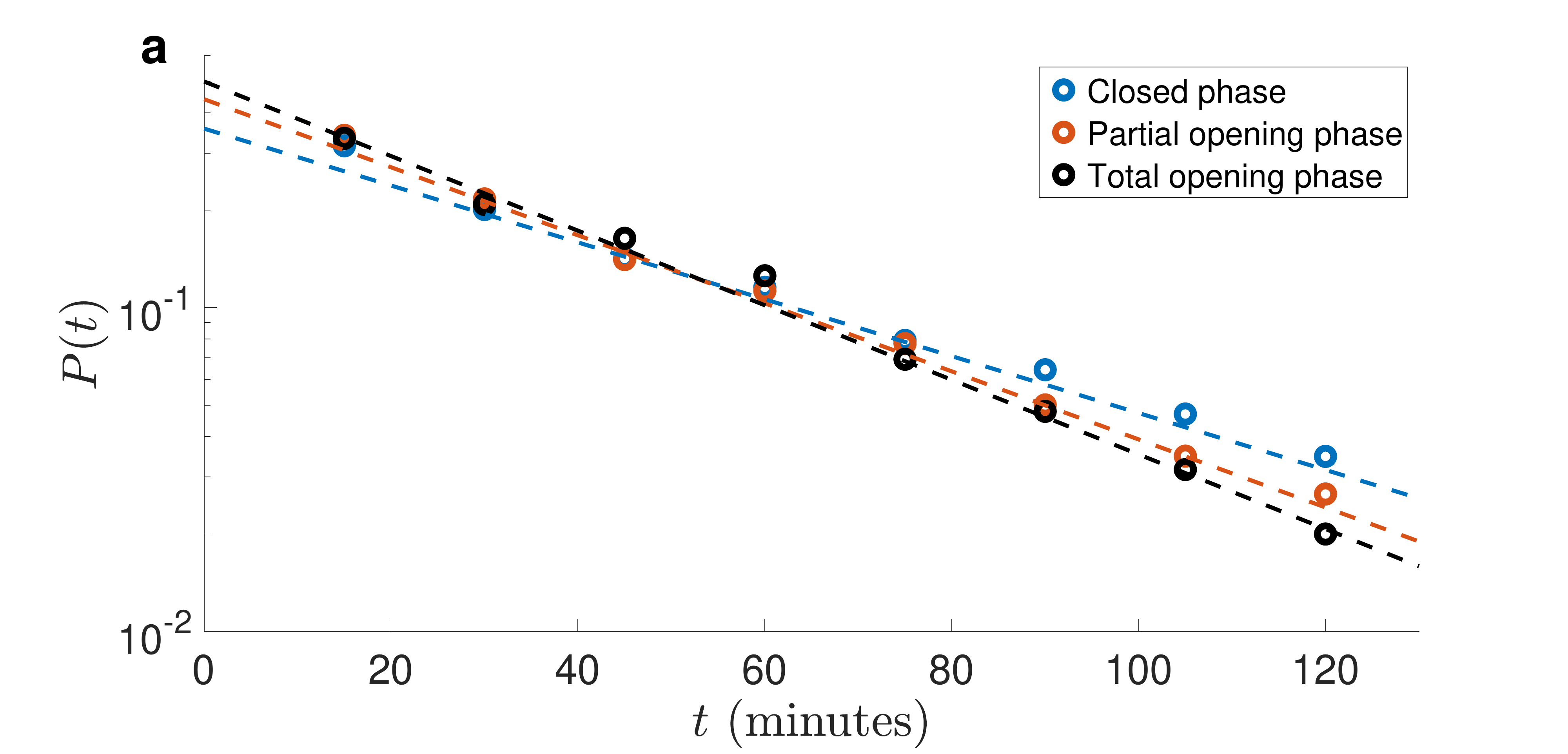}
\includegraphics[width=.75\textwidth]{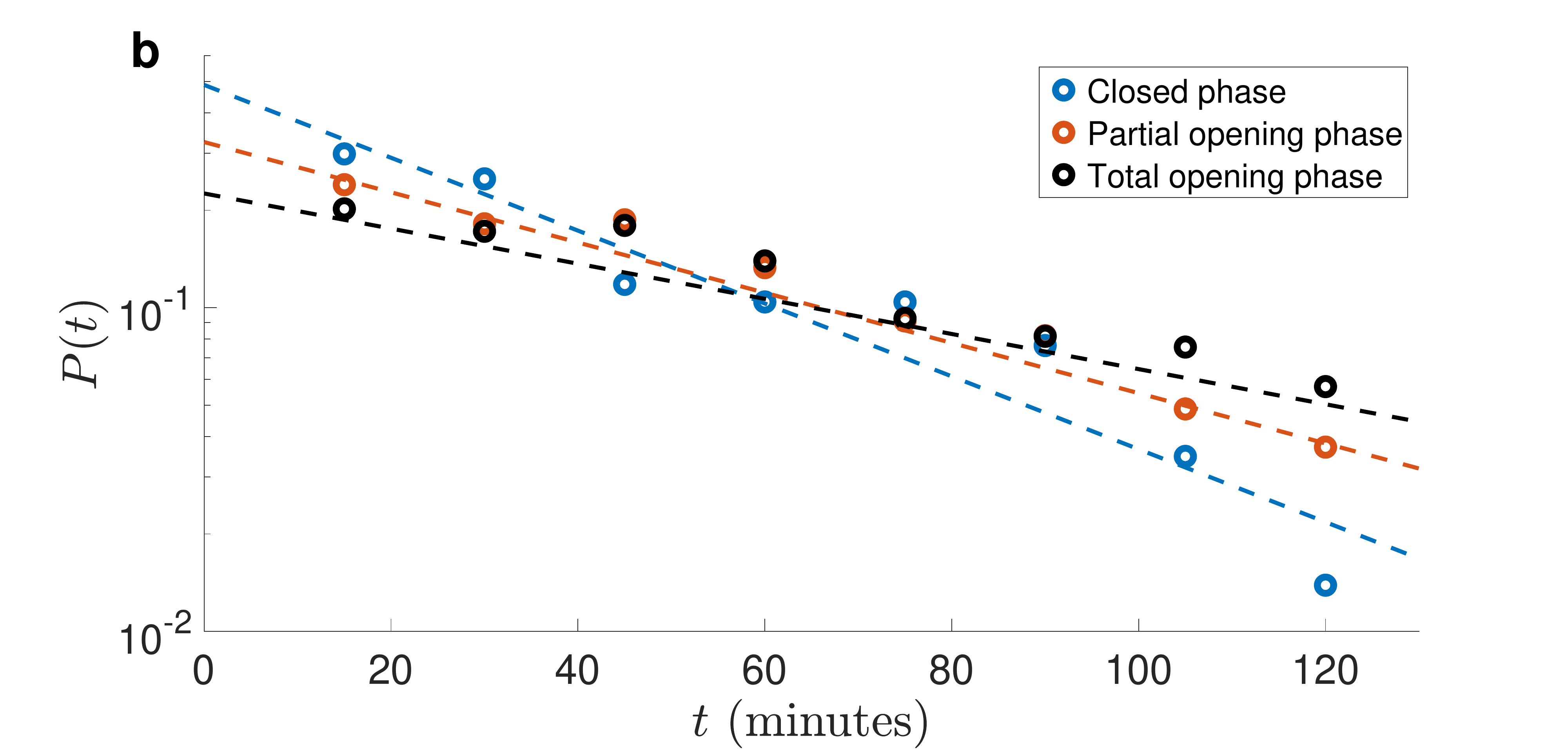}
\caption{{\bf Contact time distribution.} Probability $P(t)$ that two users are in contact for a time $t$ during working hours per day. Panel {\bf a} and {\bf b} shows the contact time distributions in the three different restriction regimes for the AP placed in atrium of Physics Department and classroom respectively. We note that all distributions have an exponential decay $f(t) \propto e^{- t/\tau}$ but with different constant $\tau$. For the contact time distributions related to the the AP installed in the atrium of Physics Department we obtain a constant $\tau_{C} = 49 \pm 2 \,m$, $\tau_{PO} = 41 \pm 1 \, m$ and $\tau_{TO} = 38 \pm 1\, m$, while for the AP placed in classroom we obtain $\tau_{C} = 39 \pm 5 \, m$, $\tau_{PO} = 56 \pm 4 \, m$ and $\tau_{TO} = 80 \pm 7 \, m$.
We note that for the atrium of Physics Department, the contact time decreases as the restrictions are removed while in the classroom the contact duration becomes longer in the opening periods.}
\label{fig:contact_time}
\end{figure*}

In order to test the effectiveness of the simplex size and of the daily link measure to characterize the critical locations, in  Fig~\ref{fig:figure7} we focus on two specific APs; one placed in a classroom of the teaching building and the other in the atrium of the Physics Department. In particular, panel {\bf c} shows that in the partial opening period the two APs have a similar size distribution $P_i(s)$ (simplices of size smaller than $5$ cannot be reconstructed due to the anonymization procedure, see Appendix). In this period, the average number of links $\langle s(s-1)\rangle_i =\int ds P_i(s)s(s-1)$ at the atrium is about $50\%$ larger than in the classroom ($\langle s(s-1)\rangle_i\approx 31.7 $ v.s. $\langle s(s-1)\rangle_i\approx 19.0$). On the other hand, in the partial opening phase, the distribution of the  number of different links per day $\tilde P_i(l)$ turns out to be very different in the two APs. In particular the average number of links per days is almost 4 times larger in the atrium than in the classroom ($\langle l \rangle_i\approx 270$ vs. $\langle l \rangle_i \approx 70$). 
This implies, as expected, a much more variable behavior of the contacts in the atrium. Panels {\bf g} and {\bf h} also show that in the atrium the occupancy is almost constant during the working hours while in the classroom people assemble mainly during lessons. The data also show that people use the two areas differently in the periods we examined. In the atrium, the average link number grows from $\langle l \rangle_i \approx 73$ in the closure period to  $\langle l \rangle_i \approx 270$ reaching  $\langle l \rangle_i\approx 445$ in the total opening phase, while in the classroom we observe a much rapid growth starting from $\langle l \rangle_i \approx 0.7$ (almost no use in the closure period) to $\langle l \rangle_i \approx 70$, up to $\langle l \rangle_i \approx 2030$ in the total opening phase. The fact that different locations are used very differently in the distinct regimes of restrictions is well confirmed in Figure \ref{fig:contact_time}, where we plot the distribution of the duration of the daily contacts between pairs of individuals in the different regimes of restrictions, in the atrium and in the considered classroom. All distributions display a typical exponential decays with a characteristic timescale and, as expected, the contacts in the atrium are typically shorter than in the classroom. Moreover, as the restriction are removed, the duration of the connection in the classroom grows, as one would expect due to the larger number of lessons; however, in the atrium, the contact time becomes shorter, showing a potential problem in monitoring the gatherings in common areas.

\subsection{Optimal monitoring of spaces}
The previous example shows that the distribution of the group sizes and of the number of new links per day provide different information to identify critical areas. Therefore, to find the APs that need to be monitored first, in each phase we characterize the AP $i$ by the average number of different links per day $\langle l \rangle_i$ and also by the average number of links present in the relevant groups, that is $\langle s(s-1)\rangle_i$. In Figure \ref{tab:ranking} and Figure \ref{tab:rankingThirdPhase} we show the APs with the largest $\langle l \rangle_i$ and $\langle s(s-1)\rangle_i$ in the partial and total opening period, respectively. There are cases of APs, particularly in study rooms and common areas, which gain more than fifteen positions in the ranking according to $\langle l \rangle_i$ compared to $\langle s(s-1)\rangle_i$, reaching the positions of critical APs. 
In this perspective we notice that in the partial opening period the atrium of the Physics Department considered in the previous section is ranked 16-th according to $\langle s(s-1)\rangle_i$ and 9-th according to $\langle l \rangle_i$, while the classroom moves from 24-th to 60-th position with the link ranking. This confirms that the two quantities classify differently the University areas according to their use. For potentially dangerous locations, the formation of large groups provides important information but the relevant measure is the number of links between different users. A high number of daily links implies a continuous reshuffling of users who could trigger superspreading events. Therefore the classification by average daily links of the APs appears to be the most relevant classification for determining the critical areas of the University Campus.

\begin{figure*}[]
    \centering
    \includegraphics[width=0.9\textwidth]{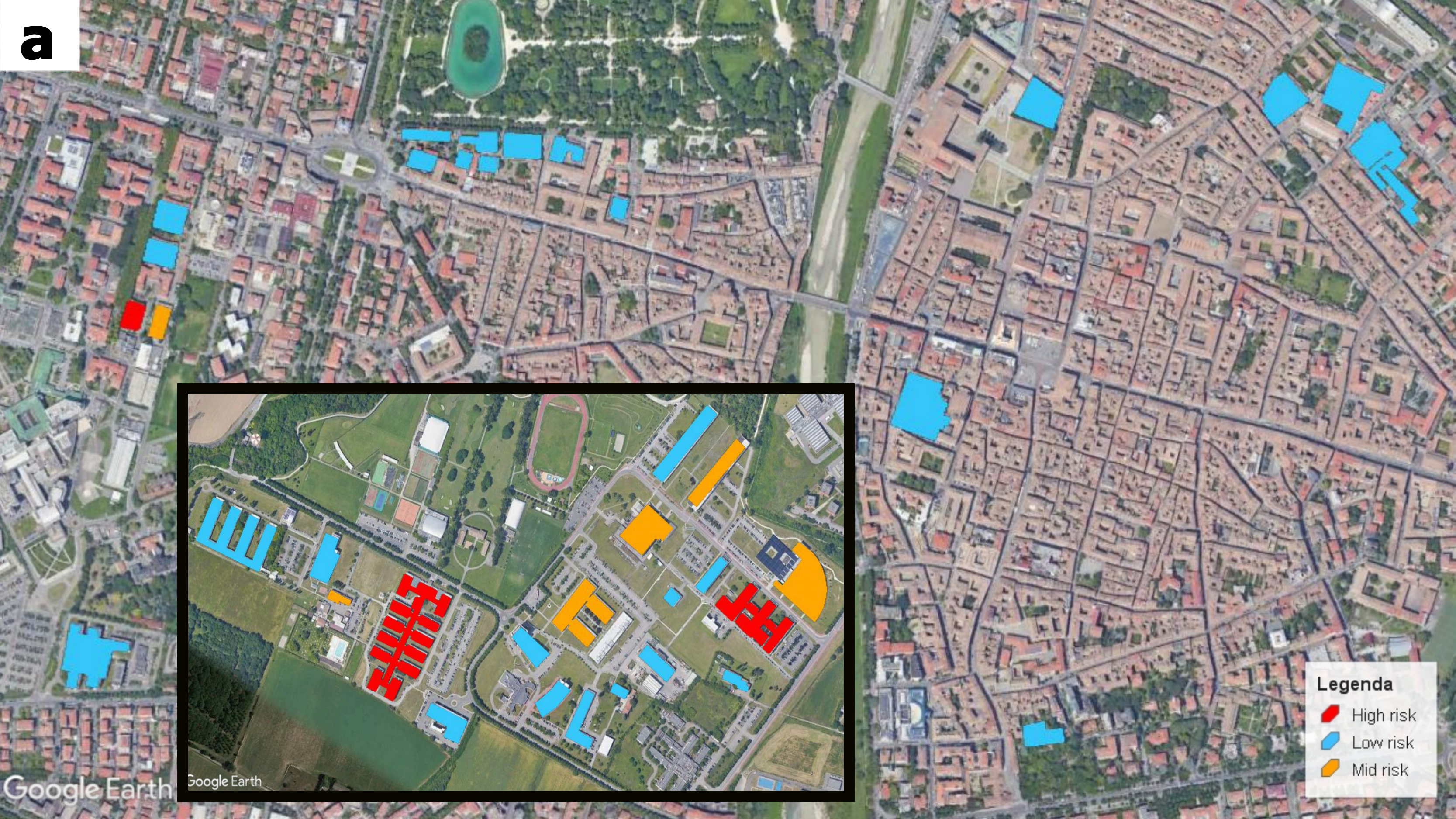}
    \includegraphics[width=0.9\textwidth]{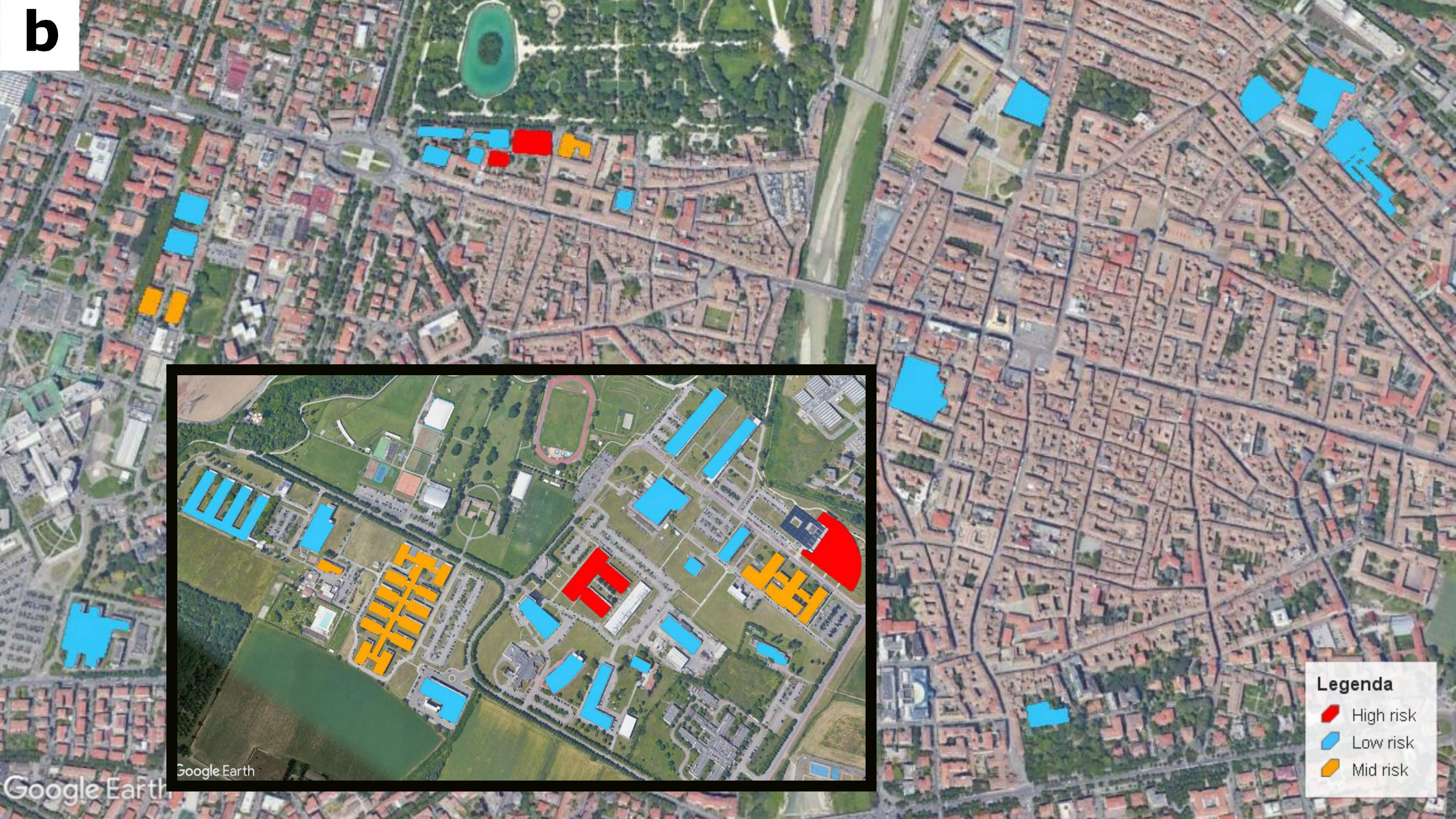}
    
    \caption{{\bf Geographical map of buildings with critical APs.} Map of Parma with the University buildings in three different colors based on the number of critical APs inside. Panel {\bf a} refers to the APs rank in Figure \ref{tab:ranking} of the partial opening period, while panel {\bf b} refers to the total opening period with the ranking of the APs given in Figure \ref{tab:rankingThirdPhase}. The light blue color represents the buildings where there are no APs in critical list, in yellow buildings where there are one or two critical APs, in red the buildings where there are more than two critical APs in the list. The return of face-to-face lessons completely changes the location of critical APs to different building, such as in the Economics department.}
    \label{fig:figure10}
\end{figure*}

Figures \ref{tab:ranking} and \ref{tab:rankingThirdPhase} also show that the location of critical AP completely changes in the partial and total opening phases. This is due to the fact that some teaching activities with a large number of students, such as in engineering and in economics, were still held completely remotely in the partial opening period. For this reason the relevant areas with typical very large attendance remained almost empty in the partial opening period. This is shown in Figure \ref{fig:figure10} where on a city map we highlight with different colors the University buildings according to the number of APs present in Figures \ref{tab:ranking} or \ref{tab:rankingThirdPhase} in the partial and total opening periods respectively. In particular we plot in light blue the buildings where there are no APs in the critical lists, in yellow buildings where there are one or two critical APs, in red the buildings where there are more than 2 critical APs in the lists.

\begin{figure*}[]
    \centering
    \includegraphics[width=0.9\textwidth]{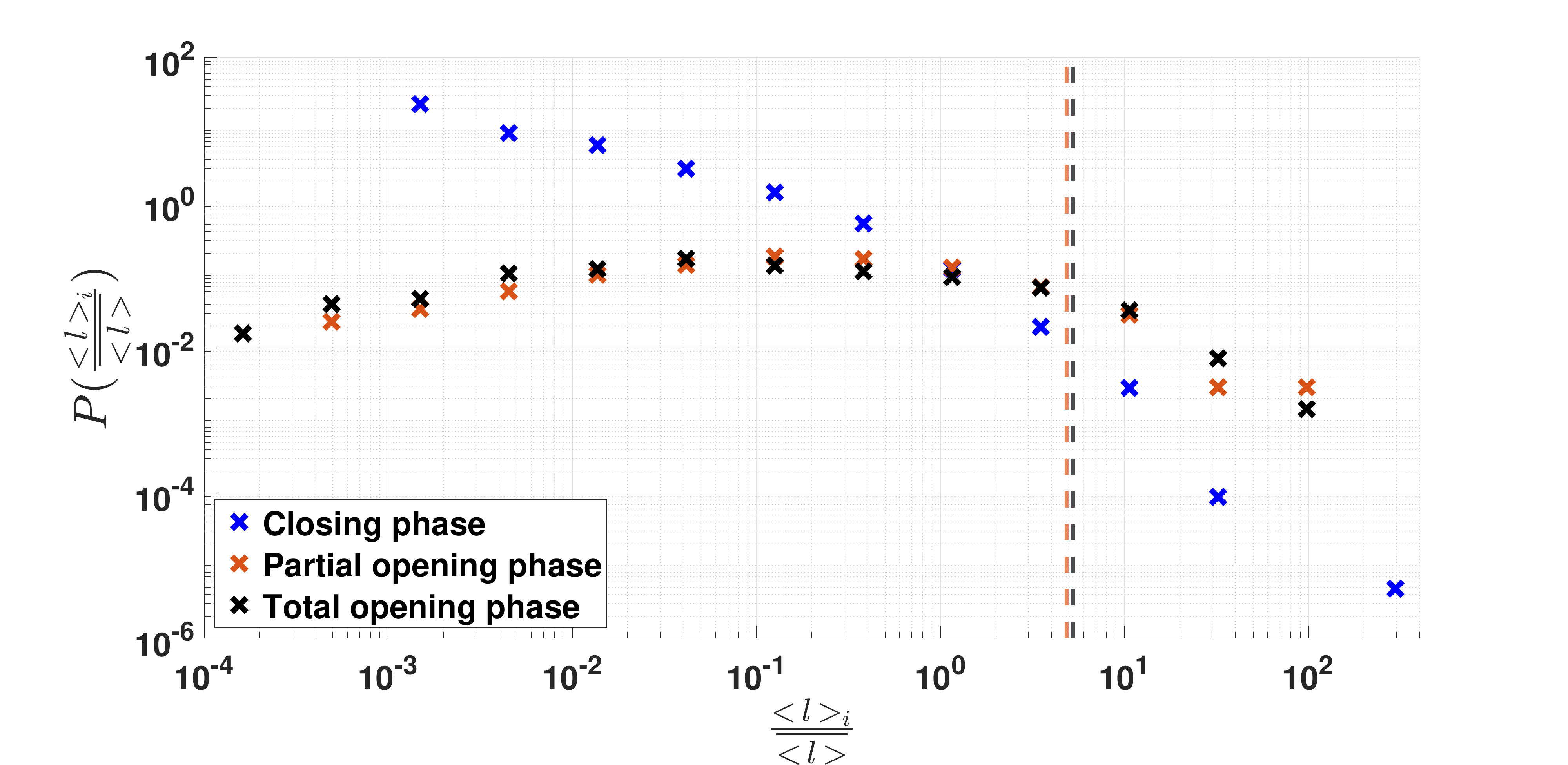}
    \includegraphics[width=0.9\textwidth]{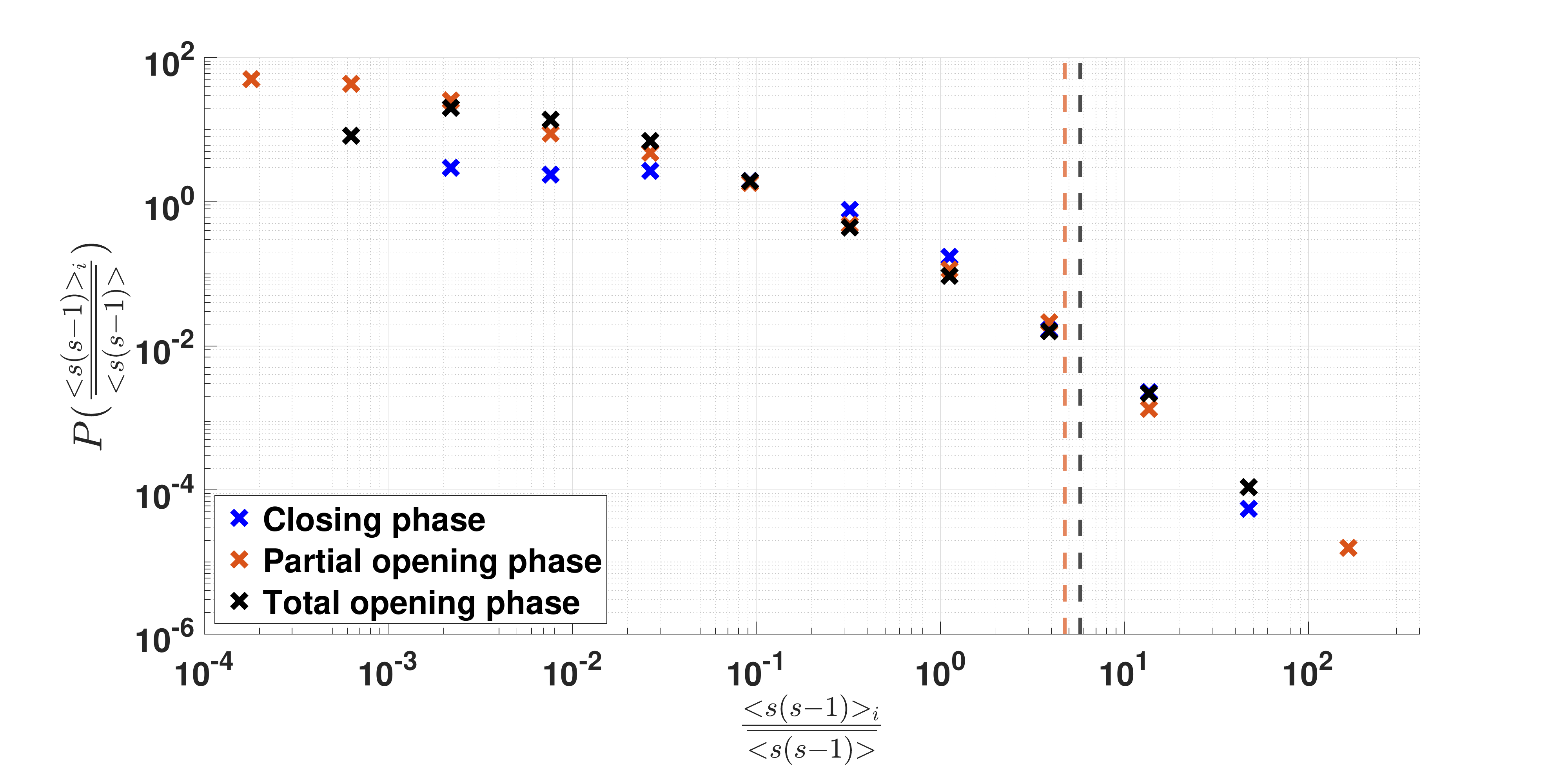}
    \caption{{\bf Links distribution among the different AP.} The plots show how the average number of different links per day $<l>_i$ and average links $<s(s-1)>_i$ are distributed among the different AP. Both quantities are normalized by their average over the different APs. The dashed lines correspond to the threshold that give the critical APs in the Figure \ref{tab:ranking} and Figure \ref{tab:rankingThirdPhase}. Panel {\bf a} shows a different distribution for the closing period compared to the opening phases, clarifying that groups are more stable in the closing phase.}
    \label{fig:figure9}
\end{figure*}

\begin{figure*}[]
    \centering
    \includegraphics[width=0.6\textwidth]{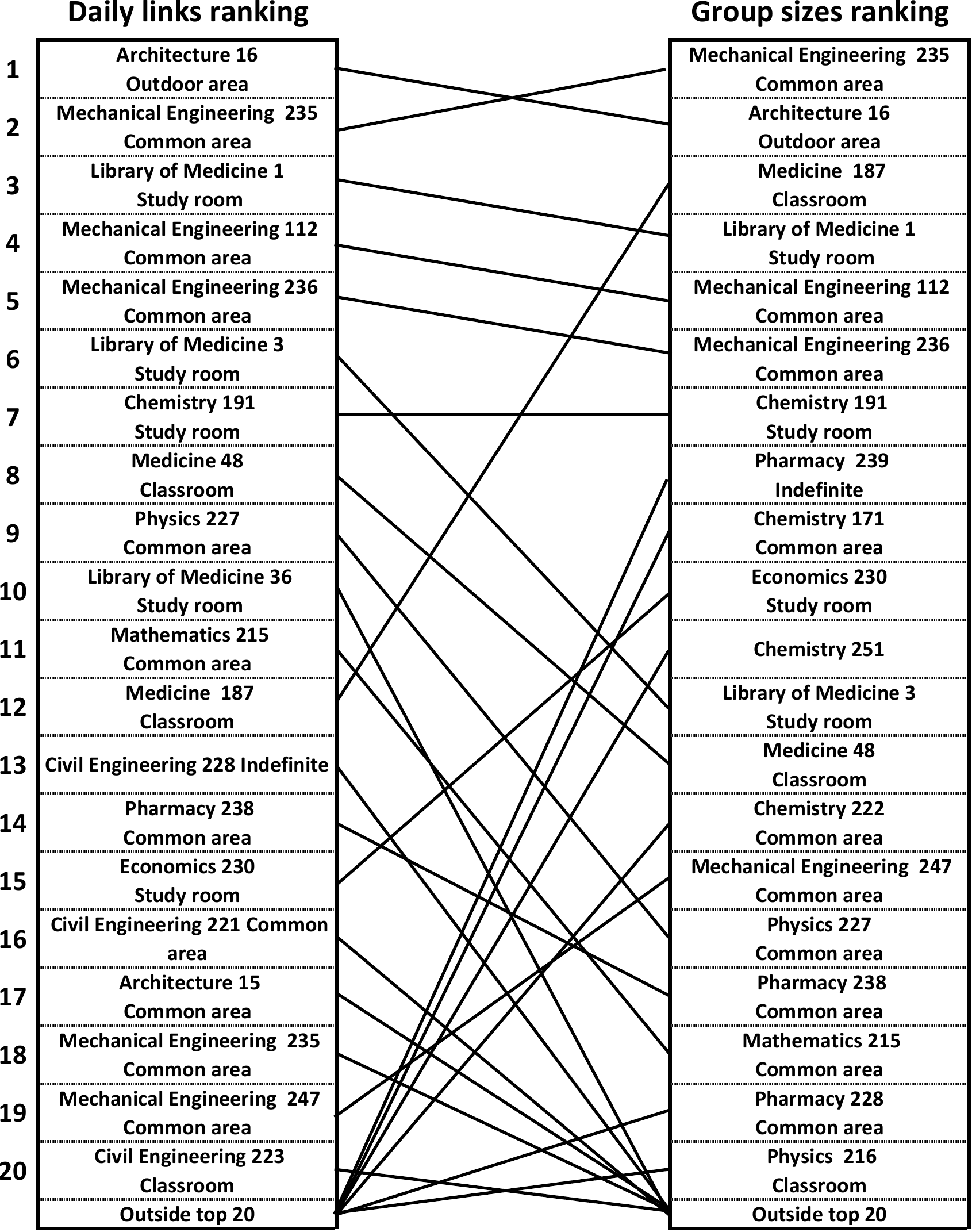}
    \caption{{\bf Rankings during partial opening phase.} Top twenty positions in the ranking by links distributions during partial opening phase compared with positions in the ranking by the size of groups. }
    \label{tab:ranking}
    
\end{figure*}

\begin{figure*}[]
    \centering
    \includegraphics[width=0.6\textwidth]{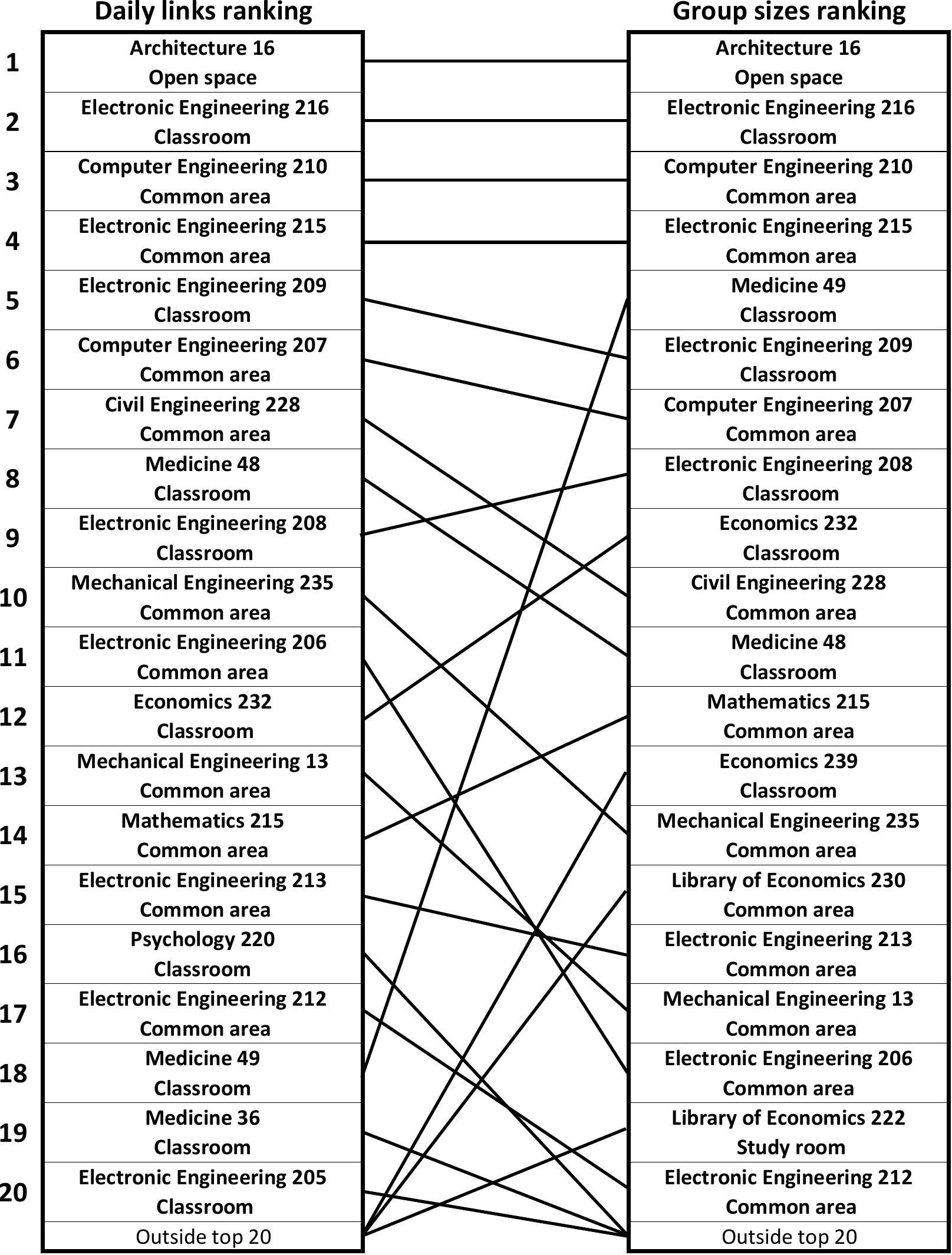}
    \caption{{\bf Rankings during total opening phase.} Top twenty positions in the ranking by links distributions during total opening phase compared with positions in the ranking by the size of groups.  }
    \label{tab:rankingThirdPhase}
    
\end{figure*}

Figure \ref{fig:figure9} shows how  $\langle l \rangle_i$ and $\langle s(s-1)\rangle_i$ are distributed among the different APs. In order to plot the data on the same scale both quantities are normalized by their average over the different APs (i.e. $\overline{\langle s(s-1)\rangle}=\frac{1}{N_{AP}}\sum^{N_{AP}}_{i=1} \langle s(s-1)\rangle_i$, $\overline{\langle l\rangle}=\frac{1}{N_{AP}}\sum^{N_{AP}}_{i=1} \langle l\rangle_i$ where $N_{AP}$ is the total  number of APs). We notice that the distributions feature a slow decays at large values, which correspond to the critical AP described in Figures \ref{tab:ranking} and \ref{tab:rankingThirdPhase}, i.e. the data above the vertical dashed lines in the figure.

%In the total opening phase, the classifications obtained have the same characteristics as the classifications of the partial opening phase, with a gain in the positions of the common areas in the classification by average daily links and a loss and the classrooms that have a loss of positions compared to the classification by average size of groups. The differences between two phases are in the location of the critical areas, with about $33\%$ of the APs being critical in both cases. In the partial opening phase the critical APs are all located in the medical department and on the campus evenly distributed among the departments, while in the total opening phase it is found that the critical APs inside of the University Campus are installed in the buildings where the lessons take place and new critical areas appear in the departments located in the city center (department of economics).

\subsection{Beyond the simplicial temporal network}
\label{sub:link_thr}
The basic reproduction number $R_0$ that we estimate in section \ref{sub:R0} has been evaluated on a temporal network model with simplices in a mean field approach without memory \cite{petri2018simplicial,mancastroppa2021sideward}. This means that it is assumed that all the $s(s-1)/2$ links formed in the group of size $s$ occur between nodes that are always different, with a total reshuffling  at each time step.
However, in our data we observe that links between the same users usually repeat at the same location (i.e. AP) and probably also at different places, leading to an overestimate of the number of contacts. Moreover, we have shown that link repetition displays a different timescales in the different restriction regimes and in the different locations. 
In this perspective, contact reshuffling may have a significant effect on epidemic spreading even in the comparison among the different phases of restrictions.

In the previous section we measured the AP link distribution. This quantity gives the number of different pairs per day formed for more than 15 minute at  each AP. In order to estimate the global effect of the contact reshuffling in the different phases, we can also consider how many different couples $\mathcal{L}_d$  are formed every day in the whole University and compare the behavior of this quantity in the different regimes. This integrated global data has been directly calculated by the ``ICT services'' office as discussed in the Appendix \ref{App:A}.
The ratio between the average number of different pairs per day in the partial opening and in closing phase is:
\begin{equation}
    \frac{\langle \mathcal{L}_d\rangle_{PO}}{\langle \mathcal{L}_d\rangle_{C}} \approx 3.70
    \label{eqn:pairsratio}
\end{equation}
We notice that the ratio obtained considering different links (Eq. \ref{eqn:pairsratio}) is significantly larger than the ratio obtained considering the total number of links given in (Eq. \ref{eqn:ratio}). This means that in the passage from partial opening to a closing period not only the size of the groups are significantly reduced but also the variability of contacts is even more suppressed. This could be due to the different behavior of students and university staff. In particular, as already observed,  in the closing phase the presence of students is reduced and probably this strongly limits the flows between the different Departments, due to face-to face lessons, and the use of common areas. On the other hand, staff members, in particular in the closing period, tend to establish contacts only with people within their own lab, displaying therefore a more stable pattern of contacts.

Finally, we measure the ratio of the average number of different links per days in the total and in the partial opening phase:
\begin{equation}
    \frac{\langle \mathcal{L}_d\rangle_{TO}}{\langle \mathcal{L}_d\rangle_{PO}} \approx 12.98
    \label{eqn:pairsRatioThirdPhase}
\end{equation}
In this case the ratio we obtain is very similar to the ratio observed between $\langle s(s-1) \rangle$ in the same phases. This seems to suggest that the effect of link reshuffling v.s. link repetition is stable in the switch between these two phases and only a global rescaling of the total number of contact is observed. We notice that now in both periods the presence of students and lessons is dominant. Therefore, the same type of space fruition takes place in the two phases in the University while, as we observed above, in the closing regime the laboratory activity provides a more stable pattern of link formation.

Our estimates for the ratio of the average number of different pairs and average number of different links per day in the two periods are robust if we change the time intervals used to define simplices. In particular, if the formation of the simplex is defined using a time interval of 5 minutes we obtain the values 3.56 and 12.50 for the ratios in Eq. \eqref{eqn:pairsratio} and \eqref{eqn:pairsRatioThirdPhase} respectively, while for a time interval of 30 minutes the results are 3.64 and 13.33.

The different behavior characterizing the closing period with respect to the partial and total opening phase is also clear in Figure \ref{fig:figure9}.  The distribution among the different APs are very similar in the two opening phases and they differs only for a global rescaling; this confirms that in the opening phases different areas are occupied in a similar way, with an analogous reshuffling mechanism and only a global shift in the number of presences. Interestingly, such regular behavior characterized by a global rescaling occurs even if the location of the critical APs drastically change in the two periods as evidenced by Figure \ref{fig:figure10}. On the other hand, in the closing period the distributions in Figure \ref{fig:figure9} are significantly different even after the global rescaling. In particular, the distribution of $\langle l \rangle_i$ shows a sharper behavior, indicating that areas characterized by a reshuffling significantly  larger than the average are very rare in the period without didactic activity.

\section{Discussion}

Data from WiFi networks provide an efficient way to monitor in (almost) real time space occupancy and mobility of users in restricted areas. These data can represent an interesting resource for maintaining a continuous monitoring in public spaces during the period of control of the epidemic that surely awaits us in the coming years. Here we used anonymized data to  
signal the most used areas of a University Campus in three different periods  with distinct containment measures. We classify the University areas according to two different quantities: the typical size of groups and the average number of links between different users formed in the area covered by the APs. Both of these quantities can be used to signal areas that are potentially dangerous due to the risk of contagion, with the second one being related to areas of intense and uncontrolled traffic that are difficult to trace and more risky. Also, these quantities can be used to efficiently monitor the different fruition of spaces from distinct class of users, such as staff and students. Within an approach to epidemic spreading on simplicial temporal networks, using the measured simplex size distribution as an input to the theory, our analysis also gives a specific estimate of the dramatic change in the reproduction number occurring in the total opening phase, due to the increasing  of contacts. \cite{mancastroppa2021sideward}.

Notably, WiFi data, further anonymized also with respect to locations, can provide a useful source to build an experimental temporal network to test specific epidemic models \cite{Swa21}. In particular, it would be very interesting  to compare the results obtained from epidemic spreading on the real network to the mean field formulation, to test the efficacy of mean field approximations.

Beyond applications to epidemic models, WiFi data represent a natural tool to obtain the structure and the evolution of groups of people that move and connect in the same spatial environment. In particular, Wi-Fi measurements allow quite naturally to detect the presence of more complex nested sub-structures (motifs) other than fully connected simplices, as evidenced in several contexts \cite{lotito22}. For these  reasons, WiFi data may be of potential interest for applications of higher order interactions models in different field such as opinion dynamics, social cooperation, complex contagion \cite{iacopini19,BATTISTON2020,BATTISTON2021}.

A further extension of our work is the analysis of other public and private spaces with a focus on public transportation there including WiFi on buses. In this context it will also be very interesting to analyze the impact of multi-modal monitoring solutions, for example combining WiFi, cellular and video-based localization. We foresee that such multi modal technologies could compensate each-other limitations enabling to analyze wide areas while limiting privacy impact.

The use of such fine-grained passive monitoring technologies could be very useful to conduct natural experiments to evaluate the impact of policies and mechanisms to steer people behaviors. For example, we plan to use this approach to evaluate the impact of COVID-related notices and signage to people crowding.

\section*{Appendix}
\subsection*{WiFi Data: preprocessing and privacy-preservation mechanisms}
\label{App:A}

All data from the unified University Wi-Fi network contained in the login management system are collected by the  ``ICT services'' office (IT). In particular, the IT  extracts every day a tabular file  (called log file) where each row represents a connection start or end event. Each line contains user information, device information and session information. In particular, to reconstruct the users' connections, the most interesting attributes are:

\begin{itemize}
    \item \textbf{Username}: user's email address;
    \item \textbf{Type of user}: this attribute allows to distinguish the users into students, structured staff and external guest;
    \item \textbf{Calling device ID}: MAC address of device to distinguish all user's devices;
    \item \textbf{Device type}: this attribute allows distinguish user's device (computers, smartphones or tablets);
    \item \textbf{Called station ID}: MAC address of the AP to which the device wants to connect;
    \item \textbf{Status type}: this attribute indicates whether this accounting request marks the beginning of the user service (\emph{Start}) or the end (\emph{Stop});
    \item \textbf{Date-time}: this attribute represent the day and the hour for the accounting request;
    \item \textbf{Session ID}: this attribute is a unique accounting ID to make it easy to match start and stop record in a log file.
    \item \textbf{Terminate cause}: this attribute indicates how the session was terminated, and can only be present in Accounting-Request records where the Status type is set to \emph{Stop}.
\end{itemize}

From these data, it is possible to reconstruct all the connection sessions to the University Wi-Fi network, their duration and their location from the APs maps.\\
\noindent
{\bf Privacy-preservation Mechanisms}. Due to the European regulation on privacy (GDPR) we cannot use directly the log files as they contain personal data \cite{10.1093/idpl/ipz026}. We conducted a Data Protection Impact Assessment (DPIA) in order to fulfill data minimization principles and be compliant with the latest Regulation on Privacy and Electronic Communications \cite{gdpr-e-privacy} Recital 25, specifically addressing WiFi monitoring, states: {\it Service providers have emerged who offer physical movements' tracking services based on the scanning of equipment related information with diverse functionalities, including ... ascertaining the number of people in a specific area ... for which the consent of end-users is not needed, provided that such counting is limited in time and space to the extent necessary for this purpose. Providers should also apply appropriate technical and organisations measures ... including pseudonymisation of the data ... erase it as soon it is not longer needed for this purpose. Providers engaged in such practices should display prominent notices located on the edge of the area of coverage informing end-users prior to entering the defined area that the technology is in operation ...}

Accordingly, we developed a procedure (script) to be run within the IT domain that completely anonymizes the data and computes the aggregated quantities described in the paper:

\begin{enumerate}

    \item IT removes all the personal data such as name and device info and runs a script which replaces the personal data with random 16-digit hexadecimal strings (pseudo-anonymizing the data). In order for the data to be consistent after being pseudo-anonymized, all correlations between attributes for all lines in log file are maintained. Every time a new personal data is replaced with a random string, IT saves the correspondence \emph{personal data-string} into \emph{keys files} and uses it every time that sensitive data is found. The seed for random string generation is changed every 24-hours.  

    \item Since in the dataset some connections are sometimes interrupted for short times due to various reason (such as weak signal or user's device in standby), if two or more consecutive connections of a single user to the same AP are present and the interval time  $\tau$  between these connections is less then 5 minutes, IT considers it a single connection with start time of first connection and stop time of last connection.

    \item IT script computes several measures that are at the core of our analysis (all data are calculated first on the whole population and then separately for staff members and students):

    \begin{itemize}
        \item {\it Group Size}: the number of devices $s_i$ connected to a given ($i^{th}$) AP for each 15 minutes. Devices must remain connected to the AP for the whole time. 
        \item {\it Presence}: the total number of individual connected to the University WIFI during each 15 minutes interval. 
        \item {\it Group Size Statistic}: the total number of 15 minutes clusters of a given size $s$ present in the whole University during each day of observation.
        \item {\it Average number of link}: the average number of links $\langle s(s-1)/2\rangle_i$ formed at the AP $i^{th}$ during each day of observation. The average is computed over the 15 minutes intervals during the considered working day.  
        \item {\it Different Link Number}: the daily number of different links $l_i$ (i.e., unique pairs of users)  connected to a given ($i^{th}$) AP for more than 15 consecutive minutes. If two users meet two or more times at the same point but at a different time, this is counted only once. 
        \item {\it Statistic of  Link duration}: the statistic of daily link duration in each access point (number of connection having a certain duration in unit of 15 minutes).
        \item {\it Total number of different links}: the total number $\mathcal{L}_d$ of different links formed daily in the whole university. If two users meet two or more times in the same day even at different APs, this is counted only once.
    \end{itemize}

\item All temporary data are deleted after 24 hours and only the above counts are retained. 

\end{enumerate}

All counters below a certain threshold (5 persons / 10 links in our implementation) are removed to avoid the potential disclosure of the presence/absence of specific individuals or links.
The resulting anonymized counts are the only data allowed to exit the IT domain for further analysis.

%%%%%%%%%%%%%%%%%%%%%%%%%%%%%%%%%%%%%%%%%%%%%%
%%                                          %%
%% Backmatter begins here                   %%
%%                                          %%
%%%%%%%%%%%%%%%%%%%%%%%%%%%%%%%%%%%%%%%%%%%%%%

%\begin{backmatter}

\section*{Data Availability Statement}
The datasets generated and analysed during the current study are not publicly available due to pending full approval from ``ICT services'' office of Parma University but can be available from Raffaella Burioni (raffaella.burioni@unipr.it) on reasonable request.

\section*{Author Contributions}
    RB AG MM and AV designed research; AG analyzed data; AG AB CV and FR collected data; RB AG MM and AV wrote the manuscript. All authors read and approved the final manuscript.

\section*{Conflict of Interest}
The authors declare that the research was conducted in the absence of any commercial or financial relationships that could be construed as a potential conflict of interest.

\section*{Acknowledgements}
  We warmly thank Marco Mancastroppa  and Andrea Prati for very useful comments and suggestions. 

\section*{Funding}
R.B., A.G. and A.V. acknowledge support from the EU-PON Project POR FSE Emilia Romagna 2018/2020 Objective 10.

%% Bibliography

% if your bibliography is in bibtex format, use those commands:
%\bibliographystyle{Frontiers-Vancouver} % Style Frontiers-Vancouver
\bibliography{bibliography}      % Bibliography file (usually '*.bib' )
% for author-year bibliography (bmc-mathphys or spbasic)
% a) write to bib file (bmc-mathphys only)
% @settings{label, options="nameyear"}
% b) uncomment next line
%\nocite{label}

%\end{backmatter}

\end{document}